\documentclass[superscriptaddress,onecolumn,pre]{revtex4}

\usepackage{amsmath}
\usepackage{graphicx,color}

\newcommand{\be}{\begin{equation}}
\newcommand{\ee}{\end{equation}}
\newcommand{\bea}{\begin{eqnarray}}
\newcommand{\eea}{\end{eqnarray}}

\begin{document}
\sloppy


\title{Brownian particles with long and short range interactions}

\author{Pierre-Henri Chavanis}
\email{chavanis@irsamc.ups-tlse.fr}
\affiliation{Laboratoire de Physique Th\'eorique (IRSAMC), CNRS and UPS, Universit\'e de Toulouse, F-31062 Toulouse, France}

\begin{abstract}

We develop a kinetic theory of Brownian particles with long and short
range interactions. Since the particles are in contact with a thermal
bath fixing the temperature $T$, they are described by the canonical
ensemble. We consider both overdamped and inertial models. In the
overdamped limit, the evolution of the spatial density is governed by
the generalized mean field Smoluchowski equation including a
mean field potential due to long-range  interactions and a generically
nonlinear barotropic pressure due to short-range interactions. This equation describes various
physical systems such as self-gravitating Brownian particles
(Smoluchowski-Poisson system), bacterial populations experiencing
chemotaxis (Keller-Segel model) and colloidal particles with capillary
interactions. We also take into account the inertia of the particles
and derive corresponding kinetic and hydrodynamic equations
generalizing the usual Kramers, Jeans, Euler and Cattaneo
equations. For each model, we provide the corresponding form of free
energy and establish the $H$-theorem and the virial theorem.  Finally,
we show that the same hydrodynamic equations are obtained in the
context of nonlinear mean field Fokker-Planck equations associated with
generalized thermodynamics. However, in that case, the nonlinear
pressure is due to the bias in the transition probabilities from one state to the other
leading to non-Boltzmannian distributions while in the former case the
distribution is Boltzmannian but the nonlinear pressure arises from the two-body
correlation function induced by the short-range potential of
interaction.  As a whole, our paper develops connections between the
topics of long-range interactions, short-range interactions, nonlinear
mean field Fokker-Planck equations and generalized thermodynamics.
It also justifies from a kinetic theory based on microscopic processes,
the basic equations that were introduced phenomenologically in gravitational
Brownian dynamics, chemotaxis and colloidal suspensions with attractive
interactions.

\end{abstract}

\maketitle

\section{Introduction}
\label{sec_introduction}

In the last ten years, the dynamics and thermodynamics of systems with long-range interactions (LRI) has been a subject of active research \cite{houches,assise,oxford,cdr}. Systems with long-range interactions are numerous in nature and concern, for example, self-gravitating systems (galaxies and globular clusters), two-dimensional turbulence (vortices and jets), non-neutral plasmas, free electron lasers (FEL), and toy models such as the Hamiltonian Mean Field (HMF) model. In these systems, the interaction potential
$u(r)$ decays with a rate slower than $1/r^d$ at large distances,
where $d$ is the dimension of space (these potentials are sometimes
called ``non-integrable''). As a result, each particle interacts with
far away particles (i.e. the
interaction is not restricted to nearest neighbors) and the energy is
{\it non-additive}. This can lead to striking properties (absent in
systems with short-range interactions) such as inequivalence of
statistical ensembles  and negative specific heats in the
microcanonical ensemble \cite{thirring,lbheat,paddy,ellis,bb,ijmpb}.

In a series of papers
\cite{paper1,paper2,paper3,paper4,paper5,angleaction}, we have
developed a general kinetic theory of systems with long-range
interactions. We have considered both isolated Hamiltonian systems
described by the microcanonical ensemble and dissipative Brownian
systems described by the canonical ensemble. In the first case, the
energy $E$ is conserved while in the latter case, the system is in
contact with a thermal bath fixing the temperature $T$.  For systems
with long-range interactions, the mean field approximation becomes
exact in a proper thermodynamic limit $N\rightarrow +\infty$ where the
coupling constant scales like $k\sim 1/N$ while the volume of the
system remains finite \cite{kac,messer,kiessling}. For Hamiltonian
systems, the dynamics of the one-body distribution function is
described, in the $N\rightarrow +\infty$ limit, by the Vlasov equation
\cite{braunhepp}. This corresponds to a collisionless regime. A more
general kinetic equation taking into account finite $N$ effects can be
obtained at the order $O(1/N)$ \cite{paper3,paper4,angleaction}. It
describes the collisional regime. For spatially homogeneous systems,
this kinetic equation corresponds to the standard Landau equation
(when collective effects are neglected) \cite{landau} or to the
Lenard-Balescu equation (when collective effects are taken into
account) \cite{lenard,balescu}. For Brownian systems, the dynamics of
the one-body distribution function is described, in the $N\rightarrow
+\infty$ limit, by the mean field Kramers equation \cite{paper2}. In
the strong friction limit $\xi\rightarrow +\infty$, it reduces to the
mean field Smoluchowski equation. In Refs. \cite{paper5,virial,hydro},
we have developed a hydrodynamics of Brownian particles in
interaction. The first two moments of the hierarchy of hydrodynamic
equations are the damped Jeans equations. If we implement a Local
Thermodynamic Equilibrium (LTE) approximation, we obtain the damped
Euler equations. The validity of this approximation has been
investigated in \cite{hydro} in the case of simple models. If we
neglect the velocity tensor in the damped Euler equations, we get a
Cattaneo-type equation for the density taking into account
memory effects.  Finally, in Ref.  \cite{paper5} we have developed a
theory of fluctuations in which the previous {\it deterministic}
partial differential equations are replaced by {\it stochastic}
partial differential equations including a multiplicative noise term
depending on position and time. Fluctuations are important (i) when
the number of particles is small, and (ii) close to a critical
point. When the system possesses metastable states, the fluctuations
can induce random transitions from one state to the other.  These
kinetic and hydrodynamic equations, involving long-range interactions,
can model various physical systems such as self-gravitating Brownian
particles
\cite{grossmann}, bacterial populations experiencing chemotaxis
\cite{ks} and colloids at a fluid interface driven by attractive
capillary interactions
\cite{colloids}. These models were initially studied in the strong
friction limit in which the dynamics of the particles is overdamped
but, later, it was realized (e.g. in chemotaxis) that inertial effects
\cite{gamba,laurencot,cschemo} and fluctuations \cite{stochemo} can
play an important role so that hydrodynamic and stochastic models have
also been introduced.

The dynamical evolution of Brownian particles in interaction is also studied in the physics of simple liquids  and colloids \cite{hansen}. In that case, the  interactions are short-ranged and the mean-field approximation is not valid. The usual approach to take correlation functions into account  is based on the Density Functional Theory (DFT) \cite{evans}  and on the Dynamical Density Functional Theory (DDFT) \cite{marconi}. The correlations between particles induced by a short-range potential of interaction lead to an excess pressure $p_{ex}({\bf r},t)$ with respect to the ideal gas pressure $p_{id}({\bf r},t)=\rho({\bf r},t) k_B T/m$. Deterministic and stochastic models  of interacting Brownian particles have been developed in that context \cite{archer}.

For certain systems, it is important to take into account both long-range and short-range interactions. For example, due to the attractive long-range interaction, self-gravitating Brownian particles, bacterial populations and colloids driven by attractive capillary interactions can collapse. In that case, the central part of the system becomes very dense. In the absence of short-range interactions, the collapse generically leads to the formation of Dirac peaks \cite{grossmann}. However, these peaks are unphysical and, in practice,  the density profile is regularized by small-scale constraints. These small-scale constraints can be due to finite size effects (the particles always have a finite size and cannot interpenetrate), steric hindrance, short-range interactions and, ultimately, quantum mechanics (Pauli exclusion principle).  These interactions will come into play when the system becomes dense enough. Their effect is to provide a nonlinear pressure that will halt the collapse and lead to a well-defined equilibrium state. An example of this regularization is provided by a gas of self-gravitating fermions in which gravitational collapse is balanced by the pressure force arising from the Pauli exclusion principle \cite{chandra,ijmpb}.

The aim of this paper is to present a unified kinetic theory of Brownian systems that takes into account {\it both} long and short range interactions. This will provide a precise justification, from a microscopic theory, of the kinetic and hydrodynamic models that have been introduced phenomenologically to describe self-gravitating Brownian particles \cite{grossmann}, chemotaxis \cite{ks} and colloids with capillary interactions \cite{colloids}. This will also make the bridge between the physics of long-range interactions \cite{cdr} and the physics of simple liquids with short-range interactions \cite{hansen,evans}. The approximations made in these two topics are radically different. For long-range interacting (LRI) systems, the mean field approximation applies  \cite{cdr} while for short-range interacting (SRI) systems, the correlations between particles are crucial and must be taken into account \cite{hansen,evans}. We shall also develop some connections with nonlinear mean field Fokker-Planck equations  (see reviews \cite{frank,nfp}) based on generalized thermodynamics \cite{tsallis,pp,tsallisbook}. In the context of generalized thermodynamics, nonlinear Fokker-Planck equations arise when the transition probabilities from one state to the other depend in a non trivial manner on the occupancy of the starting and arrival states \cite{kaniadakis}. This kinetical interaction principle (KIP) takes into account microscopic constraints that affect the dynamics of the particles at small scales. Interestingly, this yields the same type of hydrodynamic equations  \cite{gen,nfp} as in the case of short-range interactions  although the justification of the nonlinear pressure is different. In the case of generalized thermodynamics, the nonlinear pressure is due to the bias in the transition probabilities that leads to {\it non-Boltzmannian} distributions while in the DFT and DDFT used in the physics of liquids the distribution is Boltzmannian but the nonlinear pressure arises from the two-body correlation function induced by the  short-range potential of interaction. These two approaches therefore take into account microscopic constraints in a different manner. However, it is interesting to find some connections between the hydrodynamic (macroscopic) equations although the kinetic equations are different. This shows that these approaches are complementary.

As a whole, our paper develops connections between the topics of long-range interactions, short-range interactions, nonlinear mean field Fokker-Planck equations and generalized thermodynamics. The applications of our kinetic theory concern various physical systems such as self-gravitating Brownian particles, chemotaxis of bacterial populations, colloidal particles with capillary interactions, and probably others. Since our paper brings together several topics, and is addressed to an audience with different backgrounds,  it is necessary to briefly review the most important results of each topic to make the paper self-contained. However, the kinetic equations that we obtain are new and generalize those obtained {\it separately} in the physics of long-range interactions and in the physics of liquids.

\section{Statistical equilibrium state of Brownian systems}
\label{sec_equilibrium}

\subsection{The Gibbs canonical equilibrium}
\label{sec_gibbs}

We consider a system of $N$  Brownian particles with identical mass $m$ interacting via a potential $U({\bf r}_1,...,{\bf r}_N)$. We assume that the potential is of the form
\begin{equation}
U({\bf r}_1,...,{\bf r}_N)=m^2\sum_{i<j}u_{LR}(|{\bf r}_i-{\bf r}_j|)+m^2\sum_{i<j}u_{SR}(|{\bf r}_i-{\bf r}_j|)+m\sum_i\Phi_{ext}({\bf r}_i),
\label{e1}
\end{equation}
where $u_{LR}$ is a long-range binary potential, $u_{SR}$ is a short-range binary potential and $\Phi_{ext}$ is an external potential. The Hamiltonian is
\begin{equation}
H=\sum_{i=1}^{N}\frac{1}{2}m{v_{i}^{2}}+U({\bf r}_{1},...,{\bf r}_{N}),
\label{e2}
\end{equation}
where the first term is the kinetic energy $K$ and the second the potential energy $U$.
These Brownian particles are in contact with a heat bath with temperature $T$ so that they are described by the canonical ensemble. The statistical equilibrium state is given by the Gibbs canonical distribution
\begin{equation}
\label{e3} P_{N}({\bf r}_{1},{\bf v}_{1},...,{\bf r}_{N},{\bf v}_{N})={1\over
Z_{tot}(\beta)}e^{-\beta (\sum_{i=1}^{N}m{\frac{v_{i}^{2}}{2}}+U({\bf r}_{1},...,{\bf r}_{N}))},
\end{equation}
where $\beta=1/(k_B T)$ is the inverse temperature. The $N$-body distribution $P_N({\bf
r}_{1},{\bf v}_{1},...,{\bf r}_{N},{\bf v}_{N})$ gives the probability density  that the first
particle is in (${\bf r}_{1}$, ${\bf v}_{1}$), the second in (${\bf
r}_{2}$, ${\bf v}_{2}$) etc. The normalization condition $\int P_{N}d{\bf r}_{1}d{\bf v}_{1}...d{\bf r}_{N}d{\bf v}_{N}=1$ leads to the expression of the partition function: $Z_{tot}(\beta)=\int {\rm exp}\lbrack -\beta H({\bf r}_{1}, {\bf v}_{1}...,{\bf r}_{N}, {\bf v}_{N})\rbrack d{\bf r}_{1}d{\bf v}_{1}...d{\bf r}_{N}d{\bf v}_{N}$. We introduce the free energy functional
\begin{equation}
\label{ff1}F_{tot}[P_N]=E_{tot}[P_N]-TS_{tot}[P_N],
\end{equation}
where
\begin{equation}
\label{e4} S_{tot}=-k_B\int P_{N}\ln P_{N}\, d{\bf r}_{1}d{\bf v}_{1}...d{\bf r}_{N}d{\bf v}_{N},
\end{equation}
is the entropy and
\begin{eqnarray}
\label{e5}
E_{tot}=\langle H\rangle = \int P_{N}H \, d{\bf r}_{1}d{\bf v}_{1}...d{\bf r}_{N}d{\bf v}_{N},
\end{eqnarray}
is the average energy. The canonical $N$-body distribution (\ref{e3}) minimizes the free energy $F_{tot}[P_N]$ at fixed normalization. Furthermore, the value of the free energy at equilibrium, obtained by substituting the Gibbs distribution (\ref{e3}) in Eqs. (\ref{e4}) and (\ref{e5}), is $F_{tot}(\beta)=-(1/ \beta)\ln Z_{tot}(\beta)$. The average energy at equilibrium is
$E_{tot}(\beta)=-\partial\ln Z_{tot}/\partial\beta$. The fluctuations of energy are $\langle (\Delta H)^2\rangle=C/(k_B \beta^2)$ where $C=\partial E_{tot}/\partial T=-k_B\beta^2\partial E_{tot}/\partial\beta$ is the specific heat. This relation shows that the specific heat is always positive in the canonical ensemble \cite{huang}.

From Eq. (\ref{e3}), we see that the velocity dependence of the $N$-body distribution is Gaussian. Therefore, the average kinetic energy is $\langle K\rangle=dNk_BT/2$ (where $d$ is the dimension of space)
just like in a non-interacting gas.
In the following, we shall mainly focus on the configurational part of the distribution function
\begin{equation}
\label{e6} P_{N}({\bf r}_{1},...,{\bf r}_{N})={1\over
Z(\beta)}e^{-\beta U({\bf r}_{1},...,{\bf r}_{N})},
\end{equation}
which contains the non-trivial information on the system. The normalization condition $\int P_{N}d{\bf r}_{1}...d{\bf r}_{N}=1$ leads to the expression of the partition function: $Z(\beta)=\int {\rm exp}\lbrack -\beta U({\bf r}_{1},...,{\bf r}_{N})\rbrack d{\bf r}_{1}...d{\bf r}_{N}$. Due to the Gaussian
nature of the velocity distribution, we have
\begin{equation}
\label{e7} P_{N}({\bf r}_{1},{\bf v}_{1},...,{\bf r}_{N},{\bf v}_{N})=\left (\frac{\beta m}{2\pi}\right )^{dN/2} e^{-\beta\sum_{i=1}^{N}m{\frac{v_{i}^{2}}{2}}} P_{N}({\bf r}_{1},...,{\bf r}_{N}).
\end{equation}
Comparing Eqs. (\ref{e3}), (\ref{e6}) and (\ref{e7}), we find that $Z_{tot}(\beta)=(2\pi/\beta m)^{dN/2}Z(\beta)$. We introduce the configurational free energy
\begin{equation}
\label{ff2}F[P_N]=E[P_N]-TS[P_N],
\end{equation}
where
\begin{equation}
\label{e8} S=-k_B\int P_{N}\ln P_{N}\, d{\bf r}_{1}...d{\bf r}_{N},
\end{equation}
is the configurational entropy and
\begin{eqnarray}
\label{e9}
E=\langle U\rangle= \int P_{N}U\, d{\bf r}_{1}...d{\bf r}_{N},
\end{eqnarray}
is the average potential energy.
The canonical $N$-body distribution (\ref{e6}) minimizes $F[P_N]$ at fixed normalization. Furthermore, the value of the free energy at equilibrium, obtained by substituting Eq. (\ref{e6}) in Eqs. (\ref{e8}) and (\ref{e9}), is $F(\beta)=-(1/ \beta)\ln Z(\beta)$.  The average energy and the fluctuations of energy are given by expressions similar to those given above.

{\it Remark:} the free energy $F[P_N]$ does not always have a minimum. This is the case in particular for the gravitational interaction in $d=3$ due to the phenomenon of {\it gravitational collapse}. The strict statistical equilibrium state of a self-gravitating gas in the canonical ensemble is a Dirac peak containing all the particles \cite{kiesslingdirac,aaiso,post}. Such a configuration makes the free energy diverge to $-\infty$ due to the (algebraic) divergence
of the potential energy that  cannot be compensated by the (logarithmic)  divergence of the entropy in the other direction \cite{aaiso}.  However, there can exist metastable states in the form of gaseous configurations that have very long lifetimes, scaling like $e^N$ \cite{metastable}. These metastable states are local minima of the mean field free energy functional
$F[f]$ defined by Eq. (\ref{e29}). See Appendix \ref{sec_comments} for some comments about the importance of metastable states.

\subsection{The Yvon-Born-Green (YBG) hierarchy} \label{sec_ybg}

We introduce the reduced
probability distributions
\begin{equation}
\label{e10}
P_{j}({\bf r}_{1},...,{\bf r}_{j})=\int P_{N}({\bf r}_{1},...,{\bf r}_{N})\, d{\bf r}_{j+1}...d{\bf r}_{N}.
\end{equation}
Differentiating the defining relation (\ref{e10}) for $P_j$ and using Eq. (\ref{e6}), we obtain
 the YBG  hierarchy of equations \cite{hansen,paper1}:
\begin{eqnarray}
\label{e11}
{\partial P_{j}\over\partial {\bf r}_{1}}({\bf r}_{1},...,{\bf r}_{j})=-\beta m^{2} P_{j}({\bf r}_{1},...,{\bf r}_{j})\sum_{i=2}^{j} {\partial u_{1,i}\over\partial {\bf r}_{1}}\nonumber\\
-\beta m^{2} (N-j)\int P_{j+1}({\bf r}_{1},...,{\bf r}_{j+1})
{\partial u_{1,j+1}\over\partial {\bf r}_{1}}d{\bf r}_{j+1}-\beta m P_{j}({\bf r}_{1},...,{\bf r}_{j})\frac{\partial \Phi_{ext}}{\partial {\bf r}_1},
\end{eqnarray}
where $u=u_{LR}+u_{SR}$ denotes the total binary potential of interaction and we
have noted $u_{i,j}$ for $u(|{\bf r}_i-{\bf r}_j|)$. The first equation of the hierarchy is
\begin{eqnarray}
\label{e12} {\partial P_{1}\over\partial {\bf r}_{1}}({\bf r}_{1})=-\beta m^{2} (N-1) \int P_{2}({\bf r}_{1},{\bf r}_{2}){\partial u_{1,2}\over\partial {\bf r}_{1}}d{\bf r}_{2}-\beta m P_{1}({\bf r}_{1})\frac{\partial \Phi_{ext}}{\partial {\bf r}_1}.
\end{eqnarray}
If we introduce the local density $\rho({\bf r})=NmP_1({\bf r})$ and the two-body distribution function $\rho_2({\bf r},{\bf r}')=N(N-1)m^2P_2({\bf r},{\bf r}')$, the first equation of the YBG hierarchy becomes
\begin{eqnarray}
\label{e13}\frac{k_B T}{m} \nabla\rho({\bf r})=-\int \rho_{2}({\bf r},{\bf r}')\nabla u(|{\bf r}-{\bf r}'|)\, d{\bf r}'-\rho({\bf r})\nabla\Phi_{ext}({\bf r}),
\end{eqnarray}
where we used the fact that the particles are identical. This equation determines the equilibrium density profile $\rho({\bf r})$ when the two-body correlation function is known. Since the velocity distribution is Gaussian, the equilibrium distribution function $f({\bf r},{\bf v})=NmP_1({\bf r},{\bf v})$ is given by
\begin{equation}
\label{e24} f({\bf r},{\bf v})=\biggl ({\beta m\over 2\pi}\biggr )^{d/2}\rho({\bf r})e^{-\beta m{v^{2}\over 2}}.
\end{equation}
On the other hand, the free energy functionals (\ref{ff1}) and (\ref{ff2}) defined in Sec. \ref{sec_gibbs} can be written
\begin{equation}
\label{e13c} F_{tot}[P_N]={1\over 2}\int f v^{2} d{\bf r} d{\bf
v}+\frac{1}{2}\int \rho_2({\bf r},{\bf r}') u(|{\bf r}-{\bf r}'|)\, d{\bf r}d{\bf r}'+\int \rho\Phi_{ext}\, d{\bf r}+k_B T\int P_N\ln P_N\, d{\bf r}_1 d{\bf v}_1...d{\bf r}_N d{\bf v}_N,
\end{equation}
and
\begin{equation}
\label{e13b} F[P_N]=\frac{1}{2}\int \rho_2({\bf r},{\bf r}') u(|{\bf r}-{\bf r}'|)\, d{\bf r}d{\bf r}'+\int \rho\Phi_{ext}\, d{\bf r}+k_B T\int P_N\ln P_N\, d{\bf r}_1...d{\bf r}_N.
\end{equation}

\subsection{The virial theorem} \label{sec_ecano}

Taking the scalar product of Eq. (\ref{e13}) with ${\bf r}$, integrating over the entire domain and integrating by parts (assuming that the boundary terms can be neglected), we obtain the exact virial theorem
\begin{eqnarray}
\label{e14}d N k_B T+V_{LR}+V_{SR}+V_{ext}=0,
\end{eqnarray}
where
\begin{eqnarray}
\label{e15}V_{LR}=-\int \rho_{2}({\bf r},{\bf r}'){\bf r}\cdot \nabla u_{LR}({\bf r},{\bf r}')\, d{\bf r}d{\bf r}',
\end{eqnarray}
\begin{eqnarray}
\label{e16}V_{SR}=-\int \rho_{2}({\bf r},{\bf r}'){\bf r}\cdot \nabla u_{SR}({\bf r},{\bf r}')\, d{\bf r}d{\bf r}',
\end{eqnarray}
\begin{eqnarray}
\label{e17}V_{ext}=-\int \rho({\bf r}){\bf r}\cdot \nabla\Phi_{ext}({\bf r})\, d{\bf r}.
\end{eqnarray}
The first term is twice the average kinetic energy $\langle K\rangle$. The second term is the virial of the long-range interaction,  the  third term is the virial of the short-range interaction and the fourth term is the virial of the external force.

\subsection{Long-range interactions: mean field approximation}
\label{sec_elr}

We first consider a purely long-range interaction.   In that case, it has been established rigorously \cite{messer} that, in a proper thermodynamic limit $N\rightarrow +\infty$ \footnote{The usual thermodynamic limit $N\rightarrow +\infty$ with $N/V$ fixed is not relevant for systems with long-range interactions  that are generically spatially inhomogeneous, and it must be reconsidered. If we write the potential of
interaction as $u(|{\bf r}-{\bf r}'|)=k\tilde{u}(|{\bf r}-{\bf r}'|)$
where $k$ is the coupling constant, then the appropriate thermodynamic
limit for long-range interactions corresponds to $N\rightarrow
+\infty$ in such a way that the coupling constant $k\sim
1/N\rightarrow 0$ while  the volume of the system remains fixed: $V\sim 1$. This is called the Kac prescription \cite{kac}. In that limit, we have an
{\it extensive} scaling of the energy $E\sim N$ and of the entropy $S\sim N$
(while the temperature $T\sim 1$ is intensive), but the system remains
fundamentally {\it non-additive} \cite{cdr}.}, the mean field approximation is {\it exact}: the $N$-body distribution function is a product of $N$ one-body distribution functions
\begin{equation}
\label{e18} P_{N}({\bf r}_{1},...,{\bf r}_{N})=P_{1}({\bf r}_{1})...P_{1}({\bf r}_{N}).
\end{equation}
More precisely, it can be shown that the non trivial correlation functions  $P_{n}'$ of order $n$ scale like $N^{-(n-1)}$ \cite{paper1}. In particular,  $P_{2}({\bf r}_{1},{\bf r}_{2})=P_{1}({\bf r}_{1})P_{1}({\bf r}_{2})+O(1/N)$. Therefore, we can make an expansion of the equations of the YBG hierarchy in powers of the small parameter $1/N$ \cite{paper1}. Here, we limit ourselves to the limit $N\rightarrow +\infty$ so that
\begin{equation}
\label{e19} \rho_2({\bf r},{\bf r}')=\rho({\bf r})\rho({\bf r}').
\end{equation}
In the mean field approximation, the first equation (\ref{e13}) of the YBG hierarchy becomes
\begin{equation}
\label{e20}
\frac{k_B T}{m}\nabla\rho({\bf r})=-\rho({\bf r})\nabla\int\rho({\bf r}')u(|{\bf r}-{\bf r}'|)\, d{\bf r}'.
\end{equation}
After integration, this can be written in the form of a mean field Boltzmann distribution
\begin{equation}
\label{e21} \rho({\bf r})=Ae^{-\beta m\Phi({\bf r})},
\end{equation}
where
\begin{equation}
\label{e22} \Phi({\bf r})=\int \rho({\bf r}')u_{LR}(|{\bf r}-{\bf r}'|)\, d{\bf r}',
\end{equation}
is the mean field potential produced self-consistently by the smooth distribution
of particles. Therefore, the equilibrium density profile of the particles is
determined by an {\it integrodifferential} equation (\ref{e20}).
Since the velocity distribution is Gaussian, the distribution function in phase space is given by the mean-field
Maxwell-Boltzmann distribution
\begin{equation}
\label{e23} f({\bf r},{\bf v})=A'e^{-\beta m\lbrack {v^{2}\over 2}+\Phi({\bf r})\rbrack}.
\end{equation}
Equation (\ref{e20})  can be written as a condition of mean field hydrostatic equilibrium
\begin{equation}
\label{e25}
\nabla p({\bf r})=-\rho({\bf r})\nabla\Phi({\bf r}),
\end{equation}
with an equation of state
\begin{equation}
\label{e26}
p({\bf r})=\rho({\bf r})\frac{k_B T}{m}.
\end{equation}
Therefore, in the mean field approximation, the local equation of state of a Brownian gas coincides with the isothermal equation of state. Note that the pressure appearing in Eq. (\ref{e25}) is the kinetic pressure defined by
\begin{equation}
\label{e27}
p_{kin}({\bf r})=\frac{1}{d}\int f v^2\, d{\bf v}.
\end{equation}
The virial theorem is given by Eq. (\ref{e14}) where $V_{SR}=V_{ext}=0$ and where  $V_{LR}$ is replaced by  the mean field virial
\begin{equation}
\label{e28}
V=-\int\rho {\bf r}\cdot \nabla\Phi\, d{\bf r}.
\end{equation}
For the gravitational potential in $d$ dimensions, which is solution of the Poisson equation $\Delta\Phi=S_d G\rho$, the mean field virial is $V=(d-2)W$ for $d\neq 2$ (where $W=\frac{1}{2}\int \rho\Phi\, d{\bf r}$ is the mean field potential energy) and $V=-GM^2/2$ for $d=2$ \cite{virial}.
Substituting Eq. (\ref{e18}) in Eq. (\ref{e13b}), and introducing the mean field potential (\ref{e22}), we obtain the free energy functional
\begin{equation}
\label{e29} F[\rho]=E[\rho]-TS[\rho]={1\over 2}\int \rho\Phi d{\bf r}+k_B T\int {\rho\over m}\ln {\rho\over m}d{\bf r},
\end{equation}
up to an additive constant term $-Nk_B T\ln N$.
Similarly, in the mean field approximation, the free energy (\ref{e13c}) reduces to the form
\begin{equation}
\label{e32} F[f]=E[f]-TS[f]={1\over 2}\int f v^{2} d{\bf r} d{\bf
v}+ {1\over 2}\int \rho\Phi d{\bf r}+k_B T\int \frac{f}{m}\ln \frac{f}{m} d{\bf r}
d{\bf v},
\end{equation}
up to an additive constant term $-Nk_B T\ln N$. It can be shown rigorously that the equilibrium density minimizes $F[\rho]$ at fixed mass \cite{messer}. This yields the mean field Boltzmann distribution (\ref{e21}). Similarly, the equilibrium distribution function minimizes $F[f]$ at fixed mass. This yields the mean field Maxwell-Boltzmann distribution (\ref{e23}).

{\it Remark:} It has to be noted that the mean field equation (\ref{e20}) may have several solutions for a given value of mass $M$ and temperature $T$ (see, e.g. \cite{aaiso}, for self-gravitating systems). Only  (local) minima must be selected. The global minimum corresponds to the strict equilibrium state. A local (but not global) minimum corresponds to a metastable state. Saddle points must be rejected because they are ``unstable'' for some perturbations.

\subsection{Long and short-range interactions: Density Functional Theory (DFT)}
\label{sec_dft}

\subsubsection{Exact results}

We now consider a system of Brownian particles at temperature $T$ with long-range and short-range interactions (for the sake of generality we also assume that they evolve  in a fixed  external potential). A central result in the theory of fluids \cite{hansen,evans} is that, even if there exists non trivial correlations between the particles, the equilibrium density profile $\rho({\bf r})$ minimizes a free energy ${F}[\rho]$ at fixed mass. This free energy can be written as
\begin{equation}
\label{e33} {F}[\rho]=\int \rho\Phi_{ext} d{\bf r}+\frac{1}{2}\int \rho\Phi\, d{\bf r}+k_B T\int {\rho\over m}\ln {\rho\over m}d{\bf r}+F_{ex}[\rho].
\end{equation}
The first term is the potential energy associated with the external potential. The second term is the mean field potential energy associated with the long-range potential of interaction (see Sec. \ref{sec_elr}). The third term is the free energy of the ideal gas. Finally, the fourth term is the  excess free energy $F_{ex}[\rho]$. This  is a non-trivial functional determined by the short-range interactions. All the difficulty in the theory of fluids is to find some approximate forms of this functional. Once this functional  is known, the density profile, as well as all the $n$-point correlation functions, can be obtained via functional differentiation. Inversely, the excess free energy is often obtained from the study of the correlation functions. The excess free energy $F_{ex}$ is known exactly only in a few particular cases, but very good approximations can be devised in more general cases \cite{hansen}.

The fact that the density profile minimizes a free energy functional at fixed mass implies that its first constrained variations vanish. Writing $\delta F+\alpha T\delta M=0$, where $\alpha$ is a Lagrange multiplier taking into account the conservation of mass, we get
\begin{equation}
\label{e34} \frac{\delta {F}}{\delta\rho}+\alpha T=0.
\end{equation}
Taking the gradient of this expression, we obtain
\begin{equation}
\label{e35} \nabla \left (\frac{\delta {F}}{\delta\rho}\right )={\bf 0}.
\end{equation}
With the decomposition (\ref{e33}), these equations can be written
\begin{equation}
\label{e36} \frac{k_B T}{m}\lbrack 1+\ln({\rho}({\bf r})/{m})\rbrack+\Phi({\bf r})+\Phi_{ext}({\bf r})+\frac{\delta F_{ex}}{\delta\rho}+\alpha T={0},
\end{equation}
and
\begin{equation}
\label{e37} \frac{k_B T}{m}\nabla\rho+\rho\nabla \frac{\delta F_{ex}}{\delta\rho}+\rho\nabla\Phi+\rho\nabla\Phi_{ext}={\bf 0}.
\end{equation}
The equilibrium density profile is given by
\begin{equation}
\label{e36vf}
\rho({\bf r})=A e^{-\beta m \left (\Phi+\Phi_{ext}+\frac{\delta F_{ex}}{\delta\rho}\right )}.
\end{equation}
We stress, however, that the r.h.s. depends on $\rho({\bf r})$ itself, so that Eq. (\ref{e36vf}) is an integrodifferential equation.

On the other hand, the first equation (\ref{e13}) of the YBG hierarchy including long-range and short-range interactions is
\begin{eqnarray}
\label{e38} \frac{k_B T}{m}\nabla\rho=-\int \rho_{2}({\bf r},{\bf r}')\nabla u_{SR}({\bf r},{\bf r}')\, d{\bf r}'-\rho\nabla\Phi-\rho\nabla\Phi_{ext}.
\end{eqnarray}
Comparing Eqs. (\ref{e37}) and (\ref{e38}), we obtain
\begin{eqnarray}
\label{e39} \int \rho_{2}({\bf r},{\bf r}')\nabla u_{SR}({\bf r},{\bf r}')\, d{\bf r}'=\rho({\bf r})\nabla \frac{\delta F_{ex}}{\delta\rho}[\rho({\bf r})].
\end{eqnarray}
This relation is  {\it exact} at statistical equilibrium  and is a central result in the theory of fluids \cite{hansen,evans}. It relates the two-body correlation function to the excess free energy functional. Then, Eq. (\ref{e38}) can be viewed as an integrodifferential equation determining the equilibrium density profile once the excess free energy is known.

\subsubsection{Virial theorem}

Let us introduce the DFT virial
\begin{eqnarray}
\label{e40}
V_{DFT}=-\int \rho {\bf r}\cdot \nabla \frac{\delta F_{ex}}{\delta\rho}\, d{\bf r}.
\end{eqnarray}
At equilibrium, according to the exact identity (\ref{e39}), we have
\begin{eqnarray}
\label{e41}
V_{SR}=V_{DFT}.
\end{eqnarray}
On the other hand, for the long-range interaction, we can make the mean field approximation and use
\begin{eqnarray}
\label{e42}
V_{LR}=V.
\end{eqnarray}
Therefore, the virial theorem is given by Eq. (\ref{e14}) with Eqs.  (\ref{e41}) and (\ref{e42}).

\subsubsection{Barotropic pressure}
\label{sec_bp}

In a fluid, the local pressure is of the form $p=p(\rho,T)$. Since the temperature $T$ is fixed in the case of Brownian particles (canonical description), the pressure is barotropic and we shall simply write $p=p(\rho)$. In principle, the excess free energy $F_{ex}[\rho]$ can depend on the gradients of the density. This is particularly important for a fluid close to an interface \cite{evans}. Here, we shall assume that the density varies on a distance that is large with respect to the range of intermolecular forces. This is the case if the density distribution is mainly due to long-range interactions, as we shall assume in the following. With this assumption, the free energy is of the form (see Appendix \ref{sec_sim}):
\begin{equation}
\label{e43} {F}[\rho]=\frac{1}{2}\int \rho\Phi\, d{\bf r}+\int \rho\Phi_{ext} d{\bf r}+\int \rho\int^{\rho}\frac{p(\rho_1)}{\rho_1^2}\, d\rho_1\, d{\bf r}.
\end{equation}
The excess free energy is therefore
\begin{equation}
\label{e44} {F}_{ex}[\rho]=\int \rho\int^{\rho}\frac{p(\rho_1)}{\rho_1^2}\, d\rho_1\, d{\bf r}-k_B T\int {\rho\over m}\ln {\rho\over m}d{\bf r}.
\end{equation}
We note the relation
\begin{eqnarray}
\label{e45}
\nabla p(\rho)=\frac{k_{B}T}{m}\nabla\rho+\rho\nabla\frac{\delta F_{ex}}{\delta\rho}=\nabla p_{id}+\nabla p_{ex},
\end{eqnarray}
where $p_{id}({\bf r})=\rho({\bf r}) k_B T/m$ is the ideal pressure law and $p_{ex}$ is the excess pressure due to short-range interactions.
This relation can be used to determine the equation of state $p(\rho)$ corresponding to the excess free energy $F_{ex}[\rho]$ and {\it vice versa}. For an ideal fluid ($F_{ex}=0)$, we recover the perfect gas law $p({\bf r})=\rho({\bf r}) k_B T/m$. On the other hand, using the exact identity (\ref{e39}),  we find that
\begin{eqnarray}
\label{e46} \nabla p=\frac{k_B T}{m}\nabla\rho({\bf r})+\int \rho_{2}({\bf r},{\bf r}')\nabla u_{SR}({\bf r},{\bf r}')\, d{\bf r}'.
\end{eqnarray}
This equation relates the local pressure $p({\bf r})$, hence the equation of state $p(\rho)$, to the two-body correlation function. Taking the scalar product of this relation with ${\bf r}$, integrating over the entire domain and integrating by parts (assuming that boundary terms can be neglected), we obtain
\begin{eqnarray}
\label{e47} \int  p\, d{\bf r}= N k_B T-\frac{1}{d}\int \rho_{2}({\bf r},{\bf r}'){\bf r}\cdot \nabla u_{SR}({\bf r},{\bf r}')\, d{\bf r}d{\bf r}'.
\end{eqnarray}
For a spatially uniform fluid, we recover the {\it virial equation} \cite{hansen}:
\begin{eqnarray}
\label{e48} \frac{\beta P}{n}= 1-\frac{S_d}{2d}{\beta n}\int_0^{+\infty} g(\xi)u_{SR}'(\xi) \xi^d\, d\xi,
\end{eqnarray}
where $n=\rho/m=N/V$ is the number density and $g(\xi)$ the radial correlation function defined by $\rho_2({\bf r},{\bf r}')=n^2 g(|{\bf r}-{\bf r}'|)$.

\subsubsection{Hydrostatic equilibrium}

For a free energy of the form (\ref{e43}), Eq. (\ref{e37}) reduces to the condition of hydrostatic equilibrium
\begin{equation}
\label{e49} \nabla p+\rho\nabla\Phi+\rho\nabla\Phi_{ext}={\bf 0}.
\end{equation}
Comparing this relation with the first equation of the YBG hierarchy
\begin{eqnarray}
\label{e50}\frac{k_B T}{m} \nabla\rho({\bf r})+\int \rho_{2}({\bf r},{\bf r}')\nabla u_{SR}({\bf r},{\bf r}')\, d{\bf r}'+\int \rho_{2}({\bf r},{\bf r}')\nabla u_{LR}({\bf r},{\bf r}')\, d{\bf r}'+\rho({\bf r})\nabla\Phi_{ext}({\bf r})={\bf 0},
\end{eqnarray}
we note that the long-range interactions create a mean field force $-\rho\nabla\Phi$ while the short-range interactions create an excess pressure $p_{ex}({\bf r})$ with respect to the ideal pressure law $p_{id}({\bf r})=\rho({\bf r})k_B T/m$. Long range and short range interactions have therefore a very different influence on the system. On the other hand, for a free energy of the form (\ref{e43}), the DFT virial (\ref{e40}) takes the form
\begin{eqnarray}
\label{e51}
V_{DFT}=d\int p\, d{\bf r}-dNk_BT.
\end{eqnarray}
Using Eqs. (\ref{e41}), (\ref{e42}) and (\ref{e51}), the virial theorem (\ref{e14}) can be written explicitly
\begin{eqnarray}
\label{e53}
d\int p\, d{\bf r}+V+V_{ext}=0.
\end{eqnarray}

\subsubsection{Weakly inhomogeneous systems}
\label{sec_w}

In Sec. \ref{sec_bp}, we have given a first simplified expression of the excess free energy. Here, we shall briefly mention another simplified expression that has been extensively studied in the physics of liquids (see, e.g. \cite{saam77,ramakrishnan,evans}). If the density distribution varies slowly and exhibits small departures relative to some reference density $\overline{\rho}$, we can expand the functional $F_{ex}[\rho]$ to second order in $|\rho({\bf r},r)-\overline{\rho}|\ll \overline{\rho}$, thereby obtaining
\begin{eqnarray}
\label{w1}
F_{ex}[\rho]=-\frac{1}{2}k_BT\int (\rho-\overline{\rho})({\bf r},t) c(|{\bf r}-{\bf r}'|,\overline{\rho})(\rho-\overline{\rho})({\bf r}',t)\, d{\bf r}d{\bf r}',
\end{eqnarray}
where $c({\bf r},\overline{\rho})$ denotes the Ornstein-Zernike direct correlation function in the homogeneous reference system
\begin{eqnarray}
\label{w2}
c(|{\bf r}-{\bf r}'|,\overline{\rho})=-\beta \frac{\delta^2 F_{ex}}{\delta\rho({\bf r})\delta\rho({\bf r}')}.
\end{eqnarray}
There are several methods in the physics of liquids to obtain useful approximations of the direct correlation function \cite{hansen}, hence of the functional (\ref{w1}). Interestingly, we note that the functional (\ref{w1}) has the same form (up to a shift in density \footnote{This shift in density $\rho({\bf r})-\overline{\rho}$ is similar to the one arising in the modified Newtonian model studied in \cite{yukawa}.}) as the mean field free energy functional (\ref{e29}) with Eq. (\ref{e22})  provided that we view the direct correlation function as an effective binary potential
\begin{eqnarray}
\label{w3}
c({\bf r})=-\beta u_{eff}({\bf r}).
\end{eqnarray}
This makes possible to apply the results obtained for mean field potentials to this particular situation by using the correspondence (\ref{w3}).

\section{Kinetic theory of Brownian particles in the overdamped limit}
\label{sec_overdamped}

\subsection{BBGKY-like hierarchy}
\label{sec_overbbgky}

In the overdamped limit, the  dynamics of $N$ Brownian particles in interaction is governed by the coupled stochastic equations \cite{paper2}:
\begin{equation}
\label{o1} \xi\frac{d{\bf r}_{i}}{dt}=-\frac{1}{m}\nabla_{i}U({\bf
r}_{1},...,{\bf r}_{N})+\sqrt{2D}{\bf R}_{i}(t),
\end{equation}
where ${\bf R}_{i}(t)$ is a Gaussian white noise such that
$\langle {\bf R}_{i}(t)\rangle={\bf 0}$ and $\langle
R_{i}^{\alpha}(t)R_{j}^{\beta}(t')\rangle=\delta_{ij}\delta_{\alpha\beta}\delta(t-t')$. Here, $i=1,...,N$ label the particles and $\alpha=1,...,d$ the
coordinates of space. These equations can be obtained from Eqs. (\ref{i1})-(\ref{i2})  in the strong friction limit $\xi\rightarrow +\infty$ or, equivalently,
for large times $t\gg \xi^{-1}$. In these limits, it is possible to neglect the
inertial term in Eq. (\ref{i2}) leading directly to Eq. (\ref{o1}). The diffusion coefficient in phase space $D$ is related to the friction coefficient $\xi$ and to the  temperature $T$  by the Einstein relation \cite{risken}:
\begin{eqnarray}
\label{o2}
D=\frac{\xi k_B T}{m}.
\end{eqnarray}
In terms of the mobility $\mu=1/\xi m$ and of the diffusion coefficient in physical space $D_*=D/\xi^2$, the Einstein relation takes the form $D_*=\mu k_B T$. The time evolution of the $N$-body distribution $P_{N}({\bf r}_1,...,{\bf r}_N,t)$ is governed by the $N$-body Fokker-Planck equation
\begin{equation}
\label{o3} \xi{\partial P_{N}\over\partial t}=\sum_{i=1}^{N}
{\partial\over\partial {\bf r}_{i}}\cdot \biggl\lbrack
\frac{k_B T}{m}{\partial P_{N}\over\partial {\bf r}_{i}}+\frac{1}{m}
P_{N}{\partial\over\partial {\bf r}_{i}}U({\bf r}_{1},...,{\bf
r}_{N})\biggr\rbrack.
\end{equation}
This particular Fokker-Planck equation is called the $N$-body Smoluchowski equation. The $N$-body Smoluchowski equation monotonically decreases the free energy (\ref{ff2}). Indeed, a direct calculation yields the canonical $H$-theorem:
\begin{equation}
\label{o4}\dot F=-\sum_{i=1}^{N}\int {1\over\xi m P_{N}}\biggl
(\frac{k_B T}{m}{\partial P_{N}\over\partial {\bf r}_{i}}+\frac{1}{m} P_{N}\frac{\partial U}{\partial {\bf r}_i}\biggr )^{2}d{\bf r}_{1}...d{\bf r}_{N}\le 0.
\end{equation}
For a steady state, $\dot F=0$, the term in parenthesis in Eq. (\ref{o4}), which is the diffusion current in the Smoluchowski equation (\ref{o3}), vanishes. This leads to the   Gibbs canonical distribution (\ref{e6}). Therefore, the Brownian gas described by the stochastic equations (\ref{o1}) automatically relaxes towards the Gibbs distribution (if it is normalizable).

It is easy  to derive from Eq. (\ref{o3}) the equivalent of the
BBGKY hierarchy for the reduced distribution functions (\ref{e10}). It reads \cite{paper2}:
\begin{equation}
\label{o5} \xi {\partial P_{j}\over\partial t}=\sum_{i=1}^{j}
{\partial\over\partial {\bf r}_{i}}\cdot \biggl\lbrack \frac{k_B T}{m}{\partial
P_{j}\over\partial {\bf r}_{i}}+m P_{j}\sum_{k=1,k\neq i}^{j} {\partial u_{i,k}\over\partial {\bf r}_{i}}+(N-j)m\int P_{j+1}  {\partial
u_{i,j+1}\over\partial {\bf r}_{i}}d{\bf r}_{j+1}+P_j\frac{\partial\Phi_{ext}}{\partial {\bf r}_i}\biggr\rbrack.
\end{equation}
The stationary solutions of these equations coincide with the
equations (\ref{e11}) of the YBG hierarchy.  The first equation of the
BBGKY-like hierarchy is
\begin{equation}
\label{o6} \xi {\partial P_{1}\over\partial t}=
{\partial\over\partial {\bf r}_{1}}\cdot \biggl\lbrack \frac{k_B T}{m} {\partial
P_{1}\over\partial {\bf r}_{1}}+(N-1)m\int
 P_{2}{\partial {u}_{1,2}\over\partial {\bf r}_{1}}
d{\bf r}_{2}+P_1\frac{\partial\Phi_{ext}}{\partial {\bf r}_1}\biggr\rbrack.
\end{equation}
Introducing the local density $\rho({\bf r},t)=NmP_1({\bf r},t)$ and the two-body distribution function $\rho_2({\bf r},{\bf r}',t)=N(N-1)m^2P_2({\bf r},{\bf r}',t)$, we obtain the exact Smoluchowski equation
\begin{equation}
\label{o7}  \xi\frac{\partial\rho}{\partial t}=\nabla\cdot \left\lbrack
 \frac{k_B T}{m} \nabla\rho+ \int \rho_2({\bf r},{\bf r}',t)\nabla u(|{\bf r}-{\bf r}'|)\, d{\bf r}'+\rho\nabla\Phi_{ext}\right \rbrack,
\end{equation}
where we have used the fact that the particles are identical. The steady state of this equation returns the first YBG equation (\ref{e13}).

\subsection{Long-range interactions: mean field Smoluchowski equation}
\label{sec_mfs}

For systems with long-range interactions,  the mean field approximation (\ref{e18}), extended out-of-equilibrium,  is exact when $N\rightarrow +\infty$. In particular, we have
\begin{equation}
\label{zwa} \rho_{2}({\bf r},{\bf r}',t)=\rho({\bf r},t)\rho({\bf r}',t).
\end{equation}
Substituting this relation in the first equation (\ref{o7}) of the BBGKY-like hierarchy, we obtain the mean field Smoluchowski equation
\begin{equation}
\label{o9}  \xi\frac{\partial\rho}{\partial t}=\nabla\cdot \left (
 \frac{k_B T}{m}\nabla\rho+ \rho\nabla\Phi\right ),
\end{equation}
with
\begin{equation}
\label{o10} \Phi({\bf r},t)=\int \rho({\bf r}',t)u_{LR}(|{\bf r}-{\bf r}'|)\, d{\bf r}'.
\end{equation}
The free energy associated with this equation is given by Eq. (\ref{e29}).

\subsection{Long and short-range interactions: Dynamical Density Functional Theory (DDFT)}
\label{sec_sza}

In the presence of long and short range interactions, the first equation (\ref{o7}) of the BBGKY-like hierarchy can be written
\begin{equation}
\label{o11}  \xi\frac{\partial\rho}{\partial t}=\nabla\cdot \left\lbrack
 \frac{k_B T}{m}\nabla\rho+\int \rho_2({\bf r},{\bf r}',t)\nabla u_{SR}(|{\bf r}-{\bf r}'|)\, d{\bf r}'+\rho\nabla\Phi+\rho\nabla\Phi_{ext}\right \rbrack,
\end{equation}
where we have used the mean field approximation to evaluate the long-range interaction term.
This equation is not closed since it depends on the two-body distribution function. In the dynamic density functional theory of fluids, the hierarchy is closed by making the approximation \cite{marconi}:
\begin{eqnarray}
\label{o12} \int \rho_{2}({\bf r},{\bf r}',t)\nabla u_{SR}(|{\bf r}-{\bf r}'|)\, d{\bf r}'=\rho({\bf r},t)\nabla \frac{\delta F_{ex}}{\delta\rho}[\rho({\bf r},t)],
\end{eqnarray}
where $F_{ex}[\rho]$ is the excess free energy calculated at equilibrium. This relation is exact at equilibrium (see Sec. \ref{sec_dft}) and the approximation consists in extending it out-of-equilibrium with the actual density $\rho({\bf r},t)$ calculated at each time. This closure is equivalent to assuming that the two-body dynamic correlations are the same as those in an equilibrium fluid with the same one body density profile.  Although it is not possible to ascertain the validity of this approximation in the general case, it has been observed for the systems considered that this approximation gives remarkable agreement with direct Brownian $N$-body simulations. With this approximation, Eq. (\ref{o11}) becomes
\begin{equation}
\label{o13}  \xi\frac{\partial\rho}{\partial t}=\nabla\cdot \left\lbrack
 \frac{k_B T}{m}\nabla\rho+\rho\nabla \frac{\delta F_{ex}}{\delta\rho}+\rho\nabla\Phi+\rho\nabla\Phi_{ext}\right \rbrack,
\end{equation}
which is closed. If we consider a free energy of the form (\ref{e43}), then using identity (\ref{e45}), the foregoing equation can be rewritten
\begin{equation}
\label{o14}  \xi\frac{\partial\rho}{\partial t}=\nabla\cdot \left (
 \nabla p+\rho\nabla\Phi+\rho\nabla\Phi_{ext}\right ).
\end{equation}
This is a generalized mean field Smoluchowski equation including a generically nonlinear barotropic pressure $p(\rho)$ due to short-range interactions and a mean field  potential $\Phi=u_{LR} * \rho$ (where $*$ denotes the product of convolution) due to long-range interactions. This equation, coupled with a potential of the form $\Delta\Phi-k^2\Phi=S_d G\rho$ (where $k^{-1}$ is a screening length) arises in several physical problems such as self-gravitating Brownian particles \cite{grossmann}, chemotaxis \cite{ks}, colloids with capillary interactions \cite{colloids}, etc. By combining results issued from the physics of systems with long-range interactions \cite{cdr} with those issued  from the dynamic density functional theory of fluids \cite{marconi}, we have here justified this equation from a microscopic model.

Introducing the free energy functional (\ref{e33}), we can write the generalized mean field Smoluchowski equation (\ref{o13}) in the form
\begin{eqnarray}
\label{o15}
\frac{\partial\rho}{\partial t}=\nabla\cdot \left( \frac{1}{\xi}\rho\nabla\frac{\delta {F}}{\delta\rho}\right).
\end{eqnarray}
This equation monotonically decreases the free energy functional (\ref{e33}) which plays therefore  the role of a Lyapunov functional. Indeed, a straightforward calculation leads to the  $H$-theorem appropriate to the canonical ensemble
\begin{eqnarray}
\label{o16}
\dot {F}=\int \frac{\delta F}{\delta\rho}\frac{\partial\rho}{\partial t}d{\bf r}=\int \frac{\delta F}{\delta\rho}\nabla\cdot \left ( \frac{1}{\xi}\rho\nabla\frac{\delta F}{\delta\rho}\right ) d{\bf r}=-\int\frac{1}{\xi}\rho \left (\nabla\frac{\delta F}{\delta\rho}\right )^{2} d{\bf r}\le 0.
\end{eqnarray}
For a steady state, $\dot F=0$, the last term in parenthesis vanishes so that $\delta F/\delta\rho$ is uniform. This leads to Eq. (\ref{e35}). Therefore, a density profile  $\rho({\bf r})$ is a steady state of the generalized mean field Smoluchowski  equation (\ref{o15}) iff it is a critical point of $F$ at fixed mass. Furthermore, it can be shown that a steady state is linearly dynamically stable with respect to the generalized Smoluchowski equation (\ref{o15}) iff it is a (local) minimum of $F$ at fixed mass \cite{frank,nfp}. This is consistent with the condition of thermodynamical equilibrium (see Sec. \ref{sec_dft}). If $F$ is bounded from below \footnote{This is not always the case. For example, the free energy associated with the Smoluchowski-Poisson system describing self-gravitating Brownian particles is not bounded from below \cite{aaiso}. In that case, the system can experience an isothermal collapse. However, there also exists long-lived metastable states (local minima of free energy at fixed mass) on which the system can settle \cite{grossmann}.}, we know from Lyapunov's direct method that the system will converge towards a  (local) minimum of $F$ at fixed mass $M$ for $t\rightarrow +\infty$. If several (local) minima exist (metastable states), the choice of the selected equilibrium will depend on a complicated notion of basin of attraction.

{\it Remark:} Eq. (\ref{o15}) can be justified  in a phenomenological manner from the linear thermodynamics of Onsager if we interpret it as a continuity equation $\partial_t\rho+\nabla\cdot {\bf J}=0$ with a current ${\bf J}=-(1/\xi)\nabla \frac{\delta {F}}{\delta\rho}$ proportional to the gradient of a potential $\mu({\bf r})={\delta {F}}/{\delta\rho}$ that is uniform at equilibrium (see Eq. (\ref{e34})). This is precisely the way in which this equation was initially introduced in the physics of liquids \cite{evans,dieterich}; see also \cite{frank,nfp} in a more general context.

\section{Kinetic theory of Brownian systems taking into account inertial effects}
\label{sec_i}

\subsection{BBGKY-like hierarchy}
\label{sec_ibbgky}

We now take into account inertial effects and consider $N$ Brownian particles in interaction described by the coupled stochastic equations \cite{paper2}:
\begin{eqnarray}
\label{i1}
{d{\bf r}_{i}\over dt}={\bf v}_{i},
\end{eqnarray}
\begin{eqnarray}
\label{i2}
{d{\bf v}_{i}\over dt}=-\xi{\bf v}_{i}-\frac{1}{m}\nabla_{i}U({\bf
r}_{1},...,{\bf r}_{N})+\sqrt{2D}{\bf R}_{i}(t),
\end{eqnarray}
where $-\xi {\bf v}_{i}$ is a friction force, $D$ the diffusion coefficient in phase space  and ${\bf R}_{i}(t)$ a Gaussian white noise. The diffusion coefficient and the friction force are related to each other by the Einstein formula (\ref{o2}). The system is described by the canonical ensemble where the temperature $T$
measures the strength of the stochastic force (since $D\sim T$). The
stochastic process (\ref{i1})-(\ref{i2}) extends the classical
Brownian model
\cite{risken} to the case of particles {\it in interaction}. In this context, the friction is due to the presence of
an inert gas and the stochastic force is due to classical Brownian
motion, turbulence or any other stochastic effect. The friction and
the noise can also mimic the overall influence of an external medium
(not represented) with which the particles interact. This is the
notion of ``thermal bath''. The evolution of the N-body
distribution function  $P_{N}({\bf r}_1,{\bf v}_1,...,{\bf r}_N,{\bf v}_N,t)$  is governed by the $N$-body
Fokker-Planck equation
\begin{equation}
\label{i3} {\partial P_{N}\over\partial t}+\sum_{i=1}^{N}\biggl
({\bf v}_{i}\cdot {\partial P_{N}\over\partial {\bf r}_{i}}+{\bf
F}_{i}\cdot {\partial P_{N}\over\partial {\bf v}_{i}}\biggr
)=\xi \sum_{i=1}^{N} {\partial \over\partial {\bf v}_{i}}\cdot \biggl (
\frac{k_B T}{m}{\partial P_{N}\over\partial {\bf v}_{i}}+ P_{N}{\bf
v}_{i}\biggr ),
\end{equation}
where ${\bf F}_{i}=-\frac{1}{m}\nabla_{i}U({\bf r}_{1},...,{\bf r}_{N})$ is the
force by unit of mass acting on the $i$-th particle.  This particular Fokker-Planck equation is called the $N$-body Kramers equation. For   $\xi=D=0$ it reduces to the  Liouville equation which governs the evolution of  an isolated
Hamiltonian system in the microcanonical ensemble \cite{paper2}. The $N$-body Kramers equation decreases the free energy (\ref{ff1}). Indeed, a direct calculation yields the canonical $H$-theorem:
\begin{equation}
\label{i4}\dot F=-\sum_{i=1}^{N}\int {\xi m \over P_{N}}\biggl
(\frac{k_B T}{m}{\partial P_{N}\over\partial {\bf v}_{i}}+P_{N}{\bf
v}_{i}\biggr )^{2}d{\bf r}_{1}d{\bf v}_{1}...d{\bf r}_{N}d{\bf v}_{N}\le 0.
\end{equation}
For a steady state, $\dot F=0$, the term in parenthesis in Eq. (\ref{i4}), which is the diffusion current in the Kramers equation (\ref{i3}), vanishes. Since $\partial/\partial
t=0$, the advective term (l.h.s.) in Eq. (\ref{i3}) must also vanish, independently. From these
two requirements, we find that the stationary solution of the $N$-body
Fokker-Planck equation is the Gibbs canonical distribution
(\ref{e3}). Therefore, the Brownian gas described by the stochastic equations (\ref{i1})-(\ref{i2}) automatically relaxes towards the Gibbs distribution.

It is easy to derive from Eq. (\ref{i3}) the equivalent of the  BBGKY hierarchy for the reduced distribution functions $P_j({\bf r}_1,{\bf v}_1,...,{\bf r}_j,{\bf v}_j,t)$. It reads \cite{paper2}:
\begin{eqnarray}
\label{i5} {\partial P_{j}\over\partial t}+\sum_{i=1}^{j}{\bf v}_{i}\cdot {\partial
P_{j}\over\partial {\bf r}_{i}}-m\sum_{i=1}^{j}\sum_{k=1,k\neq i}^{j} \frac{\partial u_{i,k}}{\partial {\bf r}_i}\cdot {\partial P_{j}\over \partial {\bf v}_{i}}-(N-j)m\sum_{i=1}^{j}\int \frac{\partial u_{i,j+1}}{\partial {\bf r}_i}\cdot{\partial P_{j+1}\over\partial {\bf v}_{i}}\, d{\bf r}_{j+1}d{\bf v}_{j+1}\nonumber\\
-\sum_{i=1}^{j}\frac{\partial \Phi_{ext}}{\partial {\bf r}_i}\cdot {\partial P_{j}\over \partial {\bf v}_{i}}
=\xi \sum_{i=1}^{j}{\partial\over\partial {\bf v}_{i}}\cdot \biggl (\frac{k_B T}{m} {\partial P_{j}\over\partial {\bf v}_{i}}+ P_{j}{\bf v}_{i}\biggr ).
\end{eqnarray}
In particular, the first equation of the hierarchy is
\begin{eqnarray}
\label{i6} {\partial P_{1}\over\partial t}+{\bf v}_{1}\cdot {\partial
P_{1}\over\partial {\bf r}_{1}}-(N-1)m\frac{\partial}{\partial {\bf v}_1}\cdot \int P_{2}({\bf r}_1,{\bf v}_1,{\bf r}_2,{\bf v}_2,t)\frac{\partial u_{1,2}}{\partial {\bf r}_1}\, d{\bf r}_{2}d{\bf v}_{2}-\frac{\partial \Phi_{ext}}{\partial {\bf r}_1}\cdot {\partial P_{1}\over \partial {\bf v}_{1}}\nonumber\\
=\xi {\partial\over\partial {\bf v}_{1}}\cdot \biggl (\frac{k_B T}{m} {\partial P_{1}\over\partial {\bf v}_{1}}+ P_{1}{\bf v}_{1}\biggr ).
\end{eqnarray}
Introducing the distribution function $f({\bf r},{\bf v},t)=NmP_1({\bf r},{\bf v},t)$ and the two-body distribution function $f_2({\bf r},{\bf v},{\bf r}',{\bf v}')=N(N-1)m^2 P_2({\bf r},{\bf v},{\bf r}',{\bf v}')$, we obtain the exact Kramers equation
\begin{eqnarray}
\label{i7} \frac{\partial f}{\partial t}+{\bf v}\cdot \frac{\partial
f}{\partial {\bf r}}-\frac{\partial}{\partial {\bf v}}\cdot \int f_2({\bf r},{\bf v},{\bf r}',{\bf v}',t) \nabla u(|{\bf r}-{\bf r}'|)\, d{\bf r}'d{\bf v}'- \nabla\Phi_{ext}\cdot \frac{\partial f}{\partial {\bf v}}=\xi \frac{\partial}{\partial {\bf v}}\cdot \left ( \frac{k_B T}{m}\frac{\partial f}{\partial {\bf v}}+f {\bf v}\right ),
\end{eqnarray}
where we have used the fact that the particles are identical.

\subsection{Long-range interactions: mean field Kramers equation}
\label{sec_mfk}

For systems with long-range interactions,  the mean field approximation is exact when $N\rightarrow +\infty$. In particular, we have
\begin{equation}
\label{i8} f_{2}({\bf r},{\bf v},{\bf r}',{\bf
v}',t)=f({\bf r},{\bf v},t)f({\bf r}',{\bf
v}',t).
\end{equation}
Substituting this relation in the first equation  (\ref{i7}) of the BBGKY-like hierarchy, we obtain the mean field  Kramers equation
\begin{eqnarray}
\label{i9} \frac{\partial f}{\partial t}+{\bf v}\cdot \frac{\partial
f}{\partial {\bf r}}-\nabla\Phi\cdot \frac{\partial f}{\partial {\bf v}}=\xi \frac{\partial}{\partial {\bf v}}\cdot \left ( \frac{k_B T}{m}\frac{\partial f}{\partial {\bf v}}+f {\bf v}\right ),
\end{eqnarray}
where $\Phi({\bf r},t)$ is given by Eq. (\ref{o10}). For $\xi=D=0$, we recover the Vlasov equation which describes Hamiltonian systems with long-range interactions  in the limit $N\rightarrow +\infty$ \cite{paper2}.

The mean field Kramers equation (\ref{i9}) monotonically decreases the free energy functional (\ref{e32})
which plays the role of a Lyapunov functional. Indeed, a simple calculation gives
\begin{eqnarray}
\label{i9b}
\dot F=-\int\frac{\xi}{f}\left (\frac{k_B T}{m}\frac{\partial f}{\partial {\bf v}}+f{\bf v}\right )^2\, d{\bf r}d{\bf v}\le 0.
\end{eqnarray}
For a steady state, $\dot F=0$, the term in parenthesis in Eq. (\ref{i9b}), which is the diffusion current in the mean field Kramers equation (\ref{i9}),  vanishes. Since $\partial/\partial
t=0$, the advective term (l.h.s.) in Eq. (\ref{i9}) must also vanish, independently. From these
two requirements, we find that the stationary solution of the mean field Kramers equation is the  mean-field Maxwell-Boltzmann distribution (\ref{e23}). Therefore, a distribution function $f({\bf r},{\bf v})$ is a steady state of the mean field Kramers equation iff it is a critical point of free energy (\ref{e32}) at fixed mass. Furthermore, it can be shown that a steady state is linearly dynamically stable with respect to the mean field Kramers equation (\ref{i9}) iff it is a (local) minimum of $F$ at fixed mass. This is consistent with the condition of thermodynamical equilibrium (see Sec. \ref{sec_elr}).

\subsection{Long-range and short-range interactions: Dynamical Density Functional Theory (DDFT)}

In the presence of long and short range interactions, we can simplify the exact Kramers equation (\ref{i7}) by making the approximation
\begin{eqnarray}
\label{app}
f_2({\bf r},{\bf v},{\bf r}',{\bf v}',t)=f({\bf r},{\bf v},t)f({\bf r}',{\bf v}',t)\rho_2({\bf r},{\bf r}',t)/\rho({\bf r},t)\rho({\bf r}',t).
\end{eqnarray}
This yields
\begin{eqnarray}
\frac{\partial f}{\partial t}+{\bf v}\cdot \frac{\partial
f}{\partial {\bf r}}-\frac{1}{\rho({\bf r},t)}\left\lbrack  \int \rho_2({\bf r},{\bf r}',t) \nabla u(|{\bf r}-{\bf r}'|)\, d{\bf r}'\right\rbrack\cdot \frac{\partial f}{\partial {\bf v}}- \nabla\Phi_{ext}\cdot \frac{\partial f}{\partial {\bf v}}=\xi \frac{\partial}{\partial {\bf v}}\cdot \left ( \frac{k_B T}{m}\frac{\partial f}{\partial {\bf v}}+f {\bf v}\right ).
\end{eqnarray}
Using furthermore approximation (\ref{o12}) for the short-range interaction and approximation (\ref{zwa}) for the long-range interaction, we get
\begin{eqnarray}
\label{closed}
\frac{\partial f}{\partial t}+{\bf v}\cdot \frac{\partial
f}{\partial {\bf r}}-\nabla \frac{\delta F_{ex}}{\delta\rho}\cdot \frac{\partial f}{\partial {\bf v}}- \nabla(\Phi+\Phi_{ext})\cdot \frac{\partial f}{\partial {\bf v}}=\xi \frac{\partial}{\partial {\bf v}}\cdot \left ( \frac{k_B T}{m}\frac{\partial f}{\partial {\bf v}}+f {\bf v}\right ),
\end{eqnarray}
which is closed. The steady state of this equation is
\begin{eqnarray}
f({\bf r},{\bf v})=A' e^{-\beta m \left (\frac{v^2}{2}+\Phi+\Phi_{ext}+\frac{\delta F_{ex}}{\delta\rho}\right )}.
\end{eqnarray}
The integration over the velocity returns the density distribution (\ref{e36vf}).

\section{Hydrodynamics of Brownian particles in interaction}
\label{sec_hydro}

We now develop a hydrodynamical theory of Brownian particles with long and short range interactions, generalizing the theory presented in \cite{virial,paper5,hydro}.

\subsection{Damped Jeans equations}
\label{sec_jeans}

Taking the hydrodynamic moments of the exact Kramers equation (\ref{i7}) and proceeding as in \cite{virial,paper5,hydro}, we obtain
\begin{eqnarray}
\label{h1}
\frac{\partial\rho}{\partial t}+\nabla\cdot (\rho {\bf u})=0,
\end{eqnarray}
\begin{eqnarray}
\label{h2}
\rho\left \lbrack \frac{\partial {\bf u}}{\partial t}+({\bf u}\cdot \nabla){\bf u}\right \rbrack=-\frac{\partial P_{ij}}{\partial x_{j}}-\int\rho_2({\bf r},{\bf r}',t)\nabla u(|{\bf r}-{\bf r}'|)\, d{\bf r}'-\rho\nabla\Phi_{ext}-\xi\rho {\bf u},
\end{eqnarray}
where $\rho({\bf r},t)=\int f d{\bf v}$ is the density, ${\bf u}({\bf r},t)=(1/\rho)\int f{\bf v}d{\bf v}$ is the local velocity, ${\bf w}={\bf v}-{\bf u}({\bf r},t)$ is the relative velocity and $P_{ij}=\int f w_{i}w_{j}d{\bf v}$ is the kinetic pressure tensor. We also recall that the kinetic pressure is defined by $p_{kin}({\bf r},t)=\frac{1}{d}\int f w^2\, d{\bf v}$. For $\xi=0$, and in the absence of short-range interactions, Eqs. (\ref{h1})-(\ref{h2}) reduce to the equations obtained by Maxwell in his theory of gases \cite{maxwell,maxwellvankampen} and by Jeans in the context of stellar dynamics \cite{jeans,bt}. Equations (\ref{h1})-(\ref{h2}) will be called the damped Jeans equations \cite{virial}.  Using the equation of continuity (\ref{h1}), we obtain the identity
\begin{eqnarray}
\label{h3}
\rho\left \lbrack \frac{\partial {\bf u}}{\partial t}+({\bf u}\cdot \nabla){\bf u}\right \rbrack=\frac{\partial}{\partial t}(\rho {\bf u})+\frac{\partial}{\partial x_j}(\rho u_i u_j).
\end{eqnarray}
On the other hand, proceeding as in \cite{virial}, we obtain the virial theorem
\begin{eqnarray}
\label{h4}
\frac{1}{2}\frac{d^2 I}{dt^2}+\frac{1}{2}\xi\frac{dI}{dt}=2\Theta+\Pi+V_{LR}+V_{SR}+V_{ext},
\end{eqnarray}
where $I=\int \rho r^2\, d{\bf r}$ is the moment of inertia, $\Theta=\frac{1}{2}\int \rho {\bf u}^2\, d{\bf r}$ is the macroscopic kinetic energy and $\Pi=d\int p_{kin}\, d{\bf r}$ is equal to twice the microscopic kinetic energy (we have $K=\Theta+\frac{1}{2}\Pi$ where $K=\frac{1}{2}\int f v^2\, d{\bf r}d{\bf v}$ is the kinetic energy).

\subsection{Strong friction limit: Smoluchowski equation}
\label{sec_sf}

The exact Smoluchowski equation (\ref{o7}) can be derived  from the exact Kramers equation (\ref{i7}) in the strong friction limit $\xi\rightarrow +\infty$. Considering the r.h.s. of Eq. (\ref{i7}), we note that, for  $\xi\rightarrow +\infty$, the velocity distribution is Maxwellian:
\begin{equation}
\label{h5}
f({\bf r},{\bf v},t)=\biggl ({\beta m\over 2\pi}\biggr )^{d/2}\rho({\bf r},t) e^{-\beta m {v^{2}\over 2}}+O(\xi^{-1}).
\end{equation}
This implies that ${\bf u}=O(1/\xi)$, $P_{ij}=(\rho k_{B}T/m)\delta_{ij}+O(1/\xi)$ and $p_{kin}= \rho k_B T/m+O(1/\xi)$. Therefore, to leading order in $1/\xi$, the damped Jeans equation (\ref{h2}) reduces to
\begin{eqnarray}
\label{h6}
\rho {\bf u}\simeq -\frac{1}{\xi}\left (\frac{k_{B}T}{m}\nabla\rho+\int\rho_2({\bf r},{\bf r}',t)\nabla u(|{\bf r}-{\bf r}'|)\, d{\bf r}'+\rho\nabla\Phi_{ext}\right ).
\end{eqnarray}
Inserting Eq. (\ref{h6}) in the continuity equation (\ref{h1}), we
obtain the exact Smoluchowski equation (\ref{o7}). This approach shows that, for $\xi\rightarrow +\infty$, the velocity distribution is Maxwellian
and the evolution of the spatial density $\rho({\bf r},t)$ is governed by the exact Smoluchowski equation
Eq. (\ref{o7}).  For $\xi\rightarrow +\infty$, we have $\Theta\rightarrow 0$  and $\Pi\rightarrow  dNk_B T$, so that the virial theorem (\ref{h4}) becomes
\begin{eqnarray}
\label{h7}
\frac{1}{2}\xi\frac{dI}{dt}=dNk_B T+V_{LR}+V_{SR}+V_{ext}.
\end{eqnarray}
This relation can also be obtained directly from the exact Smoluchowski equation (\ref{o7}) \cite{virial}.

\subsection{Local Thermodynamic Equilibrium (LTE) approximation: damped Euler equation}
\label{sec_euler}

The damped Jeans equation (\ref{h2}) is not closed since the pressure tensor depends on the
next order moment of the velocity. Following \cite{virial,paper5,hydro},
we propose to close the hierarchy by making a local thermodynamic equilibrium (L.T.E.) approximation:
\begin{eqnarray}
\label{h8}
f_{LTE}({\bf r},{\bf v},t)=\left (\frac{\beta m}{2\pi}\right )^{d/2}
\rho({\bf r},t) e^{-\frac{1}{2}\beta m ({\bf v}-{\bf u}({\bf r},t))^2}.
\end{eqnarray}
The distribution function (\ref{h8}) minimizes the free energy (\ref{e33}) for a given value of density $\rho({\bf r},t)$ and local velocity ${\bf u}({\bf r},t)$. With the LTE approximation, the pressure tensor
takes the form
\begin{eqnarray}
\label{h9}
P_{ij}= \rho({\bf r},t)\frac{k_{B}T}{m}\delta_{ij}.
\end{eqnarray}
The kinetic pressure is $p_{kin}=\rho k_BT/m$. Substituting this result in Eq. (\ref{h2}), we obtain the damped Euler equation
\begin{eqnarray}
\label{h10}
\rho\left \lbrack \frac{\partial {\bf u}}{\partial t}+({\bf u}\cdot \nabla){\bf u}\right \rbrack=-\frac{k_{B}T}{m}\nabla\rho-\int\rho_2({\bf r},{\bf r}',t)\nabla u(|{\bf r}-{\bf r}'|)\, d{\bf r}'-\rho\nabla\Phi_{ext}-\xi\rho {\bf u}.
\end{eqnarray}
Using  $\Pi=dNk_BT$, the virial theorem (\ref{h4}) takes the form
\begin{eqnarray}
\label{h12}
\frac{1}{2}\frac{d^2 I}{dt^2}+\frac{1}{2}\xi\frac{dI}{dt}=2\Theta+dNk_BT+V_{LR}+V_{SR}+V_{ext}.
\end{eqnarray}
For $\xi\rightarrow +\infty$, we can formally neglect the advective term (l.h.s.) in Eq. (\ref{h10}) and we obtain
\begin{eqnarray}
\label{h11}
\rho {\bf u}\simeq -\frac{1}{\xi}\left (\frac{k_{B}T}{m}\nabla\rho+\int\rho_2({\bf r},{\bf r}',t)\nabla u(|{\bf r}-{\bf r}'|)\, d{\bf r}'+\rho\nabla\Phi_{ext}\right ).
\end{eqnarray}
Inserting Eq. (\ref{h11}) in the continuity equation (\ref{h1}), we recover the exact Smoluchowski equation (\ref{o7}). However, we stress that this cannot be considered as a derivation (even formal) of the exact Smoluchowski equation, unlike the derivation of Sec. \ref{sec_sf},  since the damped Euler equation (\ref{h10}) is heuristic. Indeed, there is no rigorous justification of the  local thermodynamic equilibrium (L.T.E.) approximation. Accordingly, it does not appear to be possible to rigorously derive the damped Euler equation (\ref{h10}) from the exact  Kramers equation (\ref{i7}).

{\it Remark:} The relevance of the LTE approximation has been studied in Ref. \cite{hydro}  in the case of very simple Brownian models. The conclusion is that the LTE approximation is reasonable and could be improved by replacing the temperature of the bath $T$ by a time dependent temperature $T(t)$.

\subsection{Cattaneo equation}
\label{sec_telegraph}

Using identity (\ref{h3}), the damped Euler equation (\ref{h10}) can be rewritten
\begin{eqnarray}
\label{h13}
\frac{\partial}{\partial t}(\rho {\bf u})+\nabla (\rho {\bf u}\otimes {\bf u})=-\frac{k_{B}T}{m}\nabla\rho-\int\rho_2({\bf r},{\bf r}',t)\nabla u(|{\bf r}-{\bf r}'|)\, d{\bf r}'-\rho\nabla\Phi_{ext}-\xi\rho {\bf u}.
\end{eqnarray}
This equation is hyperbolic. If we neglect the inertial term (l.h.s.) in Eq. (\ref{h13}) and
substitute the resulting expression for $\rho {\bf u}$ in the continuity equation (\ref{h1}), we obtain the exact Smoluchowski equation (\ref{o7}) that is parabolic (this is valid in a strong friction limit $\xi\rightarrow +\infty$ with the limitation indicated at the end of Sec. \ref{sec_euler}).  The Smoluchowski equation neglects memory effects and leads to infinite speed propagation. Following \cite{paper5,hydro}, we can obtain a simplified hyperbolic  model taking into account memory effects and having a finite speed propagation. Indeed, if we only neglect the nonlinear term $\nabla (\rho {\bf u}\otimes {\bf u})$ in Eq. (\ref{h13}), we obtain
\begin{eqnarray}
\label{h14}
\frac{\partial}{\partial t}(\rho {\bf u})=-\frac{k_{B}T}{m}\nabla\rho-\int\rho_2({\bf r},{\bf r}',t)\nabla u(|{\bf r}-{\bf r}'|)\, d{\bf r}'-\rho\nabla\Phi_{ext}-\xi\rho {\bf u}.
\end{eqnarray}
This approximation is exact in the linear regime close to equilibrium where
$|{\bf u}|\rightarrow 0$ \cite{paper5}. Taking the time derivative of
Eq. (\ref{h1}) and substituting Eq. (\ref{h14}) in the resulting
expression, we find that
\begin{eqnarray}
\label{h15}
\frac{\partial^{2}\rho}{\partial t^{2}}+\xi\frac{\partial\rho}{\partial t}
=\nabla\cdot  \left (\frac{k_{B}T}{m}\nabla\rho+\int\rho_2({\bf r},{\bf r}',t)\nabla u(|{\bf r}-{\bf r}'|)\, d{\bf r}'+\rho\nabla\Phi_{ext}\right ).
\end{eqnarray}
This equation, which is second order in time, is analogous to the Cattaneo equation \cite{cattaneo}, or to the  telegraph equation, which generalizes the diffusion equation by
introducing memory effects and a finite speed propagation. Using  $\Theta\rightarrow 0$, the virial theorem (\ref{h12}) takes the form
\begin{eqnarray}
\label{h16}
\frac{1}{2}\frac{d^2 I}{dt^2}+\frac{1}{2}\xi\frac{dI}{dt}=dNk_BT+V_{LR}+V_{SR}+V_{ext}.
\end{eqnarray}

\subsection{Long and short-range interactions}
\label{sec_gen}

We now consider the case where the potential of interaction is of the form $u=u_{LR}+u_{SR}$. We treat the long-range interaction in the mean field  approximation (\ref{zwa}) and the short-range interaction with the approximation (\ref{o12}).

(i) The damped Jeans equations (\ref{h1})-(\ref{h2}) become
\begin{eqnarray}
\label{h17}
\frac{\partial\rho}{\partial t}+\nabla\cdot (\rho {\bf u})=0,
\end{eqnarray}
\begin{eqnarray}
\label{h18}
\rho\left \lbrack \frac{\partial {\bf u}}{\partial t}+({\bf u}\cdot \nabla){\bf u}\right \rbrack=-\frac{\partial P_{ij}}{\partial x_{j}}-\rho\nabla\frac{\delta F_{ex}}{\delta\rho}-\rho\nabla\Phi-\rho\nabla\Phi_{ext}-\xi\rho {\bf u}.
\end{eqnarray}
The virial theorem is given by Eq. (\ref{h4}) where $V_{LR}$ is replaced by $V$ and $V_{SR}$ by $V_{DFT}$ defined by Eqs. (\ref{e28}) and (\ref{e40}).

(ii) In the strong friction limit $\xi\rightarrow +\infty$, we obtain
the generalized Smoluchowski equation
\begin{eqnarray}
\label{h19}
\frac{\partial\rho}{\partial t}
=\nabla\cdot  \left\lbrack\frac{1}{\xi}\left (\frac{k_{B}T}{m}\nabla\rho+\rho\nabla\frac{\delta F_{ex}}{\delta\rho}+\rho\nabla\Phi+\rho\nabla\Phi_{ext}\right )\right\rbrack.
\end{eqnarray}
The virial theorem is given by Eq. (\ref{h7}) where $V_{LR}$ is replaced by $V$ and $V_{SR}$ by $V_{DFT}$ defined by Eqs. (\ref{e28}) and (\ref{e40}). If the  free energy is of the form (\ref{e43}), then using the relation (\ref{e45}), we get
\begin{eqnarray}
\label{h20}
\frac{\partial\rho}{\partial t}
=\nabla\cdot  \left\lbrack\frac{1}{\xi}\left (\nabla p+\rho\nabla\Phi+\rho\nabla\Phi_{ext}\right )\right\rbrack.
\end{eqnarray}
The virial theorem is given by Eq. (\ref{h7}) where $V_{LR}$ is replaced by $V$ and $V_{SR}$ by $V_{DFT}$ defined by Eqs. (\ref{e28}) and (\ref{e51}). It can be written explicitly
\begin{eqnarray}
\label{h21}
\frac{1}{2}\xi\frac{dI}{dt}=d\int p\, d{\bf r}+V+V_{ext}.
\end{eqnarray}

(iii) The damped Euler equation (\ref{h10}) becomes
\begin{eqnarray}
\label{h22}
\rho\left \lbrack \frac{\partial {\bf u}}{\partial t}+({\bf u}\cdot \nabla){\bf u}\right \rbrack=-\frac{k_{B}T}{m}\nabla\rho-\rho\nabla\frac{\delta F_{ex}}{\delta\rho}-\rho\nabla\Phi-\rho\nabla\Phi_{ext}-\xi\rho {\bf u}.
\end{eqnarray}
The virial theorem is given by Eq. (\ref{h12}) where $V_{LR}$ is replaced by $V$ and $V_{SR}$ by $V_{DFT}$ defined by Eqs. (\ref{e28}) and (\ref{e40}). Note that Eq. (\ref{h22})
can be written in terms of the free energy (\ref{e33}) as
\begin{eqnarray}
\label{h23}
\rho\left \lbrack \frac{\partial {\bf u}}{\partial t}+({\bf u}\cdot \nabla){\bf u}\right \rbrack=-\rho\nabla \frac{\delta F}{\delta\rho}-\xi\rho {\bf u}.
\end{eqnarray}
This equation satisfies an $H$-theorem for the total free energy
\begin{eqnarray}
\label{h24}
F_{tot}[\rho,{\bf u}]=\int\rho \frac{{\bf u}^{2}}{2}d{\bf r}+\int \rho\Phi_{ext} d{\bf r}+\frac{1}{2}\int \rho\Phi\, d{\bf r}+k_B T\int {\rho\over m}\ln {\rho\over m}d{\bf r}+F_{ex}[\rho].
\end{eqnarray}
For purely long-range interactions, this functional can be deduced from the free energy (\ref{e32}) by using the LTE approximation (\ref{h8}). After some calculations \cite{nfp},  we get
\begin{eqnarray}
\label{h25}
\dot F_{tot}=-\int \xi\rho {\bf u}^2  d{\bf r}\le 0.
\end{eqnarray}
For a steady state, $\dot F_{tot}=0$, we obtain ${\bf u}={\bf 0}$. Then, Eq. (\ref{h23}) implies that $\delta F/\delta\rho$ is uniform leading to Eq. (\ref{e35}). Therefore, a steady state of the damped Euler equation is a critical point of free energy at fixed mass. Furthermore, it is dynamically stable iff it is a (local) minimum of $F$ at fixed mass $M$ (more precisely, the same results as those described at the end of Sec. \ref{sec_sza}  can be obtained). If the free energy is of the form (\ref{e43}), then using the relation (\ref{e45}), we can write the damped Euler equation as
\begin{eqnarray}
\label{h26}
\rho\left \lbrack \frac{\partial {\bf u}}{\partial t}+({\bf u}\cdot \nabla){\bf u}\right \rbrack=-\nabla p-\rho\nabla\Phi-\rho\nabla\Phi_{ext}-\xi\rho {\bf u}.
\end{eqnarray}
In that case, the total free energy is
\begin{eqnarray}
\label{ht1grt}
F_{tot}[\rho,{\bf u}]=\int\rho \frac{{\bf u}^{2}}{2}d{\bf r}+\int \rho\Phi_{ext} d{\bf r}+\frac{1}{2}\int \rho\Phi\, d{\bf r}+\int \rho\int^{\rho}{p(\rho')\over
\rho'^{2}} \,d \rho'd{\bf r}.
\end{eqnarray}
The virial theorem is given by Eq. (\ref{h12}) where $V_{LR}$ is replaced by $V$ and $V_{SR}$ by  $V_{DFT}$ defined by Eqs. (\ref{e28}) and (\ref{e51}). It can be written explicitly
\begin{eqnarray}
\label{h27}
\frac{1}{2}\frac{d^2 I}{dt^2}+\frac{1}{2}\xi\frac{dI}{dt}=2\Theta+d\int p\, d{\bf r}+V+V_{ext}.
\end{eqnarray}
In the strong friction limit $\xi\rightarrow +\infty$, the preceding equations reduce formally to those obtained in (ii). However, as we have already indicated at the end of Sec. \ref{sec_euler}, this is not the correct way to justify these equations since the damped Euler equation (\ref{h22}) is based on a LTE approximation that has no rigorous foundation.

(iv) The Cattaneo equation (\ref{h15}) becomes
\begin{eqnarray}
\label{h28}
\frac{\partial^{2}\rho}{\partial t^{2}}+\xi\frac{\partial\rho}{\partial t}
=\nabla\cdot  \left (\frac{k_{B}T}{m}\nabla\rho+\rho\nabla\frac{\delta F_{ex}}{\delta\rho}+\rho\nabla\Phi+\rho\nabla\Phi_{ext}\right ).
\end{eqnarray}
It can be written in  terms of the free energy (\ref{e33}) as
\begin{eqnarray}
\label{h29}
\frac{\partial^{2}\rho}{\partial t^{2}}+\xi\frac{\partial\rho}{\partial t}
=\nabla\cdot \left (\rho\nabla \frac{\delta F}{\delta\rho}\right )
\end{eqnarray}
The virial theorem is given by Eq. (\ref{h16}) where $V_{LR}$ is replaced by $V$ and $V_{SR}$ by $V_{DFT}$ defined by Eqs. (\ref{e28}) and (\ref{e40}). If the free energy is of the form (\ref{e43}), then using the relation (\ref{e45}), we can write the Cattaneo equation as
\begin{eqnarray}
\label{h30}
\frac{\partial^{2}\rho}{\partial t^{2}}+\xi\frac{\partial\rho}{\partial t}
=\nabla\cdot  \left (\nabla p+\rho\nabla\Phi+\rho\nabla\Phi_{ext}\right ).
\end{eqnarray}
The virial theorem is given by  Eq. (\ref{h16}) where $V_{LR}$ is replaced by $V$ and $V_{SR}$ by $V_{DFT}$ defined by Eqs. (\ref{e28}) and (\ref{e51}). It can be written explicitly
\begin{eqnarray}
\label{h31}
\frac{1}{2}\frac{d^2 I}{dt^2}+\frac{1}{2}\xi\frac{dI}{dt}=d\int p\, d{\bf r}+V+V_{ext}.
\end{eqnarray}

{\it Remark:} starting from Eq. (\ref{closed}), we could also derive a system of hydrodynamic equations for the density $\rho({\bf r},t)$, the local velocity  ${\bf u}({\bf r},t)$ and the kinetic temperature $T({\bf r},t)$. They are given by Eqs. (75)-(77) of \cite{hydro}, with $\Phi({\bf r})$ replaced by $\Phi({\bf r},t)+\Phi_{ext}({\bf r})+\frac{\delta F}{\delta\rho}$.

\subsection{Stochastic kinetic equations}
\label{sec_sto}

When there exists metastable states (local minima of free energy), and when the number of particles is not too large
\footnote{This is the case, for example, in chemotaxis where the number of particles
(bacteria, cells,...) can be relatively small.},
it is important to take  fluctuations into account.  In that case, the preceding deterministic
equations must be replaced by stochastic equations including a noise term. These equations involve the coarse-grained
density $\overline{\rho}({\bf r},t)$ and coarse-grained distribution function
$\overline{f}({\bf r},{\bf v},t)$ that are spatial and/or time coarse-grained averages of the density
operators $\rho_d({\bf r},t)$ and $f_d({\bf r},{\bf v},t)$, instead of the average density $\rho({\bf r},t)$ or
average distribution function ${f}({\bf r},{\bf v},t)$ (see Appendix \ref{sec_comments}). For finite $N$ systems, or when we are close to a critical point, the
fluctuations can induce random transitions from one metastable state
to the other. The
system will visit these minima randomly and pass from one state to the other (of course, the global
minimum of free energy is the most frequently visited). On the other hand, for large $N$, or when we are far from a critical point, the metastable states have extremely  long lifetimes (larger in practice than the duration of the experiment), scaling like $e^N$ \cite{metastable}, and they are as much relevant as
fully stable states. In that case, we can use the deterministic equations of the previous sections. The relation between deterministic and stochastic equations, and the importance of metastable states, are further discussed in Appendix \ref{sec_comments}.

The stochastic coarse-grained Kramers  equation is \cite{paper5}:
\begin{eqnarray}
\label{s1} \frac{\partial \overline{f}}{\partial t}+{\bf v}\cdot \frac{\partial
\overline{f}}{\partial {\bf r}}-\frac{\partial}{\partial {\bf v}}\cdot \int \overline{f_2}({\bf r},{\bf v},{\bf r}',{\bf v}',t) \nabla u(|{\bf r}-{\bf r}'|)\, d{\bf r}'d{\bf v}'=\xi \frac{\partial}{\partial {\bf v}}\cdot \left (\frac{k_B T}{m}\frac{\partial \overline{f}}{\partial {\bf v}}+\overline{f} {\bf v}\right )+\frac{\partial}{\partial {\bf v}}\cdot \left (\sqrt{2\xi k_B T\overline{f}}\, {\bf Q}({\bf r},{\bf v},t)\right ),\nonumber\\
\end{eqnarray}
where ${\bf Q}({\bf r},{\bf v},t)$ is a Gaussian white noise satisfying $\langle {\bf Q}({\bf r},{\bf v},t)\rangle={\bf 0}$ and $\langle  {\bf Q}({\bf r},{\bf v},t){\bf Q}({\bf r}',{\bf v}',t')\rangle=\delta({\bf r}-{\bf r}')\delta({\bf v}-{\bf v}')\delta(t-t')$. We can now generalize the derivation of the hydrodynamical equations given in Ref. \cite{paper5}.

The stochastic damped Jeans equations are
\begin{eqnarray}
\label{s2}
\frac{\partial\overline{\rho}}{\partial t}+\nabla\cdot (\overline{\rho} \, \overline{{\bf u}})=0,
\end{eqnarray}
\begin{eqnarray}
\label{s3}
\overline{\rho}\left \lbrack \frac{\partial \overline{\bf u}}{\partial t}+(\overline{\bf u}\cdot \nabla)\overline{\bf u}\right \rbrack=-\frac{\partial \overline{P}_{ij}}{\partial x_{j}}-\int\overline{\rho_2}({\bf r},{\bf r}',t)\nabla {u}(|{\bf r}-{\bf r}'|)\, d{\bf r}'-\overline{\rho}\nabla\Phi_{ext}-\xi\overline{\rho}\,  \overline{\bf u}-\sqrt{2\xi k_B T\overline{\rho}}\, {\bf R}({\bf r},t).
\end{eqnarray}
Using approximations similar to those made in Secs. \ref{sec_mfs} and \ref{sec_sza}, but now applying to the coarse-grained distribution functions, we replace the integral involving the correlation function by
\begin{eqnarray}
\label{s4} \int \overline{\rho_{2}}({\bf r},{\bf r}',t)\nabla u({\bf r},{\bf r}')\, d{\bf r}'=\overline{\rho}({\bf r},t)\nabla \frac{\delta F_{ex}}{\delta\overline{\rho}}[\overline{\rho}({\bf r},t)]+\overline{\rho}({\bf r},t)\nabla\overline{\Phi}({\bf r},t),
\end{eqnarray}
where $F_{ex}[\overline{\rho}]$ is the equilibrium free energy functional and $\overline{\Phi}({\bf r},t)$ the mean field potential (\ref{o10}) determined by the coarse-grained density $\overline{\rho}({\bf r},t)$. This leads to the stochastic damped Jeans equation
\begin{eqnarray}
\label{s5}
\overline{\rho}\left \lbrack \frac{\partial \overline{\bf u}}{\partial t}+(\overline{\bf u}\cdot \nabla)\overline{\bf u}\right \rbrack=-\frac{\partial \overline{P}_{ij}}{\partial x_{j}}-\overline{\rho}\nabla \frac{\delta F_{ex}}{\delta\overline{\rho}}-\overline{\rho}\nabla\overline{\Phi}-\overline{\rho}\nabla\Phi_{ext}-\xi\overline{\rho}\,  \overline{\bf u}-\sqrt{2\xi k_B T\overline{\rho}}\, {\bf R}({\bf r},t).
\end{eqnarray}

In the strong friction limit, we get the
stochastic coarse-grained Smoluchowski equation
\begin{eqnarray}
\label{s6}
\frac{\partial\overline{\rho}}{\partial t}=\nabla\cdot \left ( \frac{1}{\xi}\overline{\rho}\nabla\frac{\delta F}{\delta\overline{\rho}}\right )+\nabla\cdot \left (\sqrt{\frac{2k_{B}T\overline{\rho}}{\xi}}{\bf R}\right ),
\end{eqnarray}
where $F[\overline{\rho}]$ is the free energy functional (\ref{e33}).

If we make the LTE approximation, we get the stochastic coarse-grained damped Euler equation
\begin{eqnarray}
\label{s7}
\overline{\rho}\left \lbrack \frac{\partial \overline{\bf u}}{\partial t}+(\overline{\bf u}\cdot \nabla)\overline{\bf u}\right \rbrack=-\overline{\rho}\nabla \frac{\delta F}{\delta\overline{\rho}}-\xi\overline{\rho}\,  \overline{\bf u}-\sqrt{2\xi k_B T\overline{\rho}}\, {\bf R}({\bf r},t).
\end{eqnarray}

Finally, the stochastic coarse-grained Cattaneo equation is
\begin{eqnarray}
\label{s8}
\frac{\partial^{2}\overline{\rho}}{\partial t^{2}}+\xi\frac{\partial\overline{\rho}}{\partial t}
=\nabla\cdot \left (\overline{\rho}\nabla \frac{\delta F}{\delta\overline{\rho}}\right )+\nabla\cdot \left (\sqrt{{2\xi k_{B}T\overline{\rho}}}\, {\bf R}\right ).
\end{eqnarray}

The virial theorem associated with the stochastic damped Euler equation (\ref{s7}) is
\begin{eqnarray}
\label{s9}
\frac{1}{2}\frac{d^2 I}{dt^2}+\frac{1}{2}\xi\frac{d I}{dt}=2\Theta-\int \overline{\rho}{\bf r}\cdot \nabla\frac{\delta F}{\delta\overline{\rho}}\, d{\bf r}-\sqrt{2\xi k_B T I(t)}\eta(t),
\end{eqnarray}
where $\eta(t)$ is a Gaussian white noise satisfying $\langle \eta(t)\rangle=0$ and $\langle \eta(t)\eta(t')\rangle=\delta(t-t')$. Using Eqs. (\ref{e33}), (\ref{e28}) and (\ref{e40}), the second term in the r.h.s. can be decomposed into
\begin{eqnarray}
\label{s10}
-\int \overline{\rho}{\bf r}\cdot \nabla\frac{\delta F}{\delta\overline{\rho}}\, d{\bf r}=dNk_B T+V_{DFT}+V+V_{ext}.
\end{eqnarray}
The virial theorem associated with the  stochastic Smoluchowski equation (\ref{s6}) is
\begin{eqnarray}
\label{s11}
\frac{1}{2}\xi\frac{d I}{dt}=-\int \overline{\rho}{\bf r}\cdot \nabla\frac{\delta F}{\delta\overline{\rho}}\, d{\bf r}-\sqrt{2\xi k_B T I(t)}\eta(t),
\end{eqnarray}
and the virial theorem associated with the  stochastic Cattaneo equation (\ref{s8}) is
\begin{eqnarray}
\label{s12}
\frac{1}{2}\frac{d^2 I}{dt^2}+\frac{1}{2}\xi\frac{d I}{dt}=-\int \overline{\rho}{\bf r}\cdot \nabla\frac{\delta F}{\delta\overline{\rho}}\, d{\bf r}-\sqrt{2\xi k_B T I(t)}\eta(t).
\end{eqnarray}

\section{Nonlinear mean field Fokker-Planck  equations and generalized
thermodynamics} \label{sec_gt}

Hydrodynamic equations similar to those derived previously, including a long-range potential of interaction and a generically nonlinear equation of state $p(\rho)$  taking into account small-scale constraints, had previously been derived \cite{gen} in the context of nonlinear mean field Fokker-Planck equations (NFP) and generalized thermodynamics pioneered by Tsallis \cite{tsallis} and Plastino \& Plastino \cite{pp} (see \cite{frank,nfp,tsallisbook} for reviews). However, the origin of the nonlinear equation of state $p(\rho)$ is physically different in the two approaches. In the context of generalized thermodynamics, the equation of state arises from the non-Boltzmannian nature of the distribution function (and $p=p_{kin}$ represents the kinetic pressure) while in the theory of fluids developed in Sec. \ref{sec_hydro}, the distribution function is Boltzmannian and the equation of state arises from  the two-body correlation function induced by the short-range potential of interaction (and $p$ represents the thermodynamic pressure). Despite this fundamental difference, the hydrodynamic (macroscopic) equations have the same mathematical form! It is therefore interesting to compare these two approaches in detail in order to stress their analogies and differences.

\subsection{Overdamped model}
\label{sec_genk}

Let us consider a system of Brownian particles in interaction in the overdamped limit. We assume that the particles interact via a mean field potential $\Phi({\bf r},t)$ given by Eq. (\ref{o10}) and that they are submitted to an external potential $\Phi_{ext}({\bf r})$. For the moment, we ignore small-scale constraints. In that
case, the motion of a particle is described  by the stochastic Langevin equation:
\begin{eqnarray}
\label{go1} \xi\frac{d{\bf
r}}{dt}=-\nabla\Phi({\bf r},t)-\nabla\Phi_{ext}({\bf r})+\sqrt{2D}{\bf R}(t),
\end{eqnarray}
where $\xi$ and $D$ are the coefficients of friction and diffusion and ${\bf R}(t)$ is a Gaussian white noise. The temperature is given  by the Einstein relation $D=\xi T$ (we take the Boltzmann constant and the mass of the particles equal to unity). The corresponding Fokker-Planck equation is the mean field Smoluchowski equation (\ref{o9}).
In order to take into account microscopic constraints that affect the motion of the particles, Kaniadakis \cite{kaniadakis}  has proposed to modify the form of the transition probability  from one state to another. This kinetical interaction principle (KIP) can take into account exclusion or inclusion constraints that enhance or inhibit the transition. This can model  for example quantum effects, close packing effects, steric hindrance... Let us define the transition probability of a particle from position ${\bf r}$ to position ${\bf r}'$ by (see \cite{kaniadakis} and Sec. 2.11 of \cite{nfp} for details):
\begin{eqnarray}
\label{go2}
\pi({\bf r}\rightarrow {\bf r}')=w({\bf r},{\bf r}'-{\bf r})a\lbrack\rho({\bf r},t)\rbrack b\lbrack\rho({\bf r}',t)\rbrack,
\end{eqnarray}
where $w({\bf r},{\bf r}'-{\bf r})$ is the transition rate that only depends on the nature of the interaction between the test particle and the bath, and $a(\rho)$ and $b(\rho)$ are positive functions.
Linear kinetics corresponds to $a(\rho)=\rho$ and
$b(\rho)=1$: the transition probability is proportional to the
density of the starting state and independent on the density of the arrival  state. It leads to the ordinary Fokker-Planck equation
(\ref{o9}). Here, we assume a more general
dependence on the occupancy in the starting and arrival states. This creates a biais with respect to the ordinary situation. Using a first neighbor approximation and an  extension of the Kramers-Moyal expansion based on the transition probability (\ref{go2}), Kaniadakis \cite{kaniadakis} obtains a nonlinear Fokker-Planck equation of the form
\begin{eqnarray}
\label{go3}
\xi\frac{\partial\rho}{\partial t}=\nabla\cdot \left \lbrack T h(\rho)\nabla\rho+ g(\rho)\nabla (\Phi+\Phi_{ext})\right \rbrack,
\end{eqnarray}
where the functions $h(\rho)$ and $g(\rho)$ are related to the bias $a(\rho)$ and
$b(\rho)$ in the transition  probabilities by
\begin{eqnarray}
\label{go4}
g(\rho)=a(\rho)b(\rho), \quad h(\rho)=b(\rho)a'(\rho)-a(\rho)b'(\rho).
\end{eqnarray}
The generalized free energy associated with the NFP equation (\ref{go3}) is
\begin{eqnarray}
\label{go5}
F[\rho]=E-TS=\frac{1}{2}\int \rho\Phi\, d{\bf r}+\int \rho\Phi_{ext}\, d{\bf r}+T\int C(\rho)\, d{\bf r},
\end{eqnarray}
where
\begin{eqnarray}
\label{go6}
S=-\int C(\rho)\, d{\bf r},\qquad C''(\rho)=\frac{h(\rho)}{g(\rho)},
\end{eqnarray}
is a ``generalized entropy'' determined by the ratio of the functions $h(\rho)$ and $g(\rho)$ \cite{kaniadakis,gen,frank,nfp,curado}. We shall assume that $C$ is convex (i.e. $C''\ge 0$). In the absence of microscopic constraint, $h(\rho)=1$ and $g(\rho)=\rho$. In that case, Eq. (\ref{go3}) reduces to the mean field Smoluchowski equation (\ref{o9}), the entropy (\ref{go6}) reduces to the Boltzmann entropy $S=-\int\rho\ln\rho\, d{\bf r}$ and the free energy (\ref{go5}) reduces to the Boltzmann free energy (\ref{e29}). We note that the NFP equation (\ref{go3}) can be written in the form \cite{frank,nfp}:
\begin{eqnarray}
\label{go7}
\frac{\partial\rho}{\partial t}=\nabla\cdot \left (\frac{1}{\xi} g(\rho)\nabla\frac{\delta F}{\delta\rho}\right ).
\end{eqnarray}
This equation satisfies an  $H$-theorem:
\begin{eqnarray}
\label{go8}
\dot F=-\int \frac{1}{\xi} g(\rho)\left (\nabla\frac{\delta F}{\delta\rho}\right )^2\, d{\bf r}\le 0.
\end{eqnarray}
For a steady state, $\dot F=0$, Eq. (\ref{go8}) implies that $\delta F/\delta\rho$ is uniform. Therefore, a steady state of the NFP equation (\ref{go7})  is a critical point of free energy at fixed mass. Furthermore, it is dynamically stable iff it is a (local) minimum of $F$ at fixed mass $M$ (more precisely, the same results as those described at the end of Sec. \ref{sec_sza}  can be obtained). Writing $\delta F+\alpha T\delta M=0$, where $\alpha$ is a Lagrange multiplier, we find that the steady states of the generalized mean field Smoluchowski equation  (\ref{go3}) are given by
\begin{eqnarray}
\label{sste}
\rho({\bf r})=(C')^{-1}\left\lbrack -\beta(\Phi+\Phi_{ext})-\alpha\right \rbrack.
\end{eqnarray}

In the following, we shall assume that $g(\rho)=\rho$. In that case, the NFP equation (\ref{go3}) becomes
\begin{eqnarray}
\label{go9}
\xi\frac{\partial\rho}{\partial t}=\nabla\cdot \left \lbrack T\rho C''(\rho)\nabla\rho+ \rho\nabla (\Phi+\Phi_{ext})\right \rbrack.
\end{eqnarray}
It can be written in the form of a generalized mean field Smoluchowski  equation
\begin{eqnarray}
\label{go10}
\xi\frac{\partial\rho}{\partial t}=\nabla\cdot   \left\lbrack \nabla p+
\rho  \nabla (\Phi+\Phi_{ext}) \right\rbrack,
\end{eqnarray}
with a barotropic equation of state $p(\rho)$ given by
\begin{eqnarray}
\label{go11}
p'(\rho)=T\rho C''(\rho).
\end{eqnarray}
Equation (\ref{go10}) is mathematically equivalent to Eq. (\ref{o14}). Its steady state is given by the condition of hydrostatic equilibrium (\ref{e49}). Since $C(\rho)$ is convex, we find that $p'(\rho)\ge 0$. A first integration gives
\begin{eqnarray}
\label{go12}
p(\rho)=T\rho^{2} \left \lbrack \frac{C(\rho)}{\rho} \right\rbrack'=T\lbrack C'(\rho)\rho-C(\rho)\rbrack.
\end{eqnarray}
A second integration leads to the identity
\begin{eqnarray}
\label{go13}
TC(\rho)=\rho \int^{\rho}\frac{p(\rho_1)}{\rho_1^{2}}d\rho_1.
\end{eqnarray}
Therefore, the free energy (\ref{go5}) can be rewritten
\begin{eqnarray}
\label{go14}
F[\rho]=\frac{1}{2}\int\rho\Phi \, d{\bf r}+\int\rho\Phi_{ext}\, d{\bf r}+\int  \rho \int^{\rho}\frac{p(\rho_1)}{\rho_1^{2}}d\rho_1d{\bf r}.
\end{eqnarray}
It coincides with the free energy (\ref{e43}) describing a fluid with weak density gradients. Therefore, the two approaches (DDFT and generalized thermodynamics) lead to similar equations but for different reasons. We finally note that the NFP equation (\ref{go10}) can be derived from the generalized stochastic process
\begin{equation}
\label{go15} \xi{d{\bf r}\over dt}=-\nabla\Phi({\bf r},t)-\nabla \Phi_{ext}({\bf
r})+\sqrt{\frac{\xi p(\rho({\bf r},t))}{\rho({\bf r},t)}}{\bf R}(t),
\end{equation}
in which the noise term explicitly depends on the density of particles around the particle under consideration. This is a phenomenological manner to take into account microscopic constraints that can affect the motion of the particles \cite{borland,frank,nfp}.

\subsection{Inertial model}
\label{sec_gi}

We now extend the previous approach in phase space in order to take into account the inertia of the particles. In the absence of microscopic constraints, the mean field dynamics of Brownian particles is described by the Langevin equations
\begin{eqnarray}
\label{gi1} \frac{d{\bf r}}{dt}={\bf v},
\end{eqnarray}
\begin{eqnarray}
\label{gi2} \frac{d{\bf v}}{dt}=-\xi{\bf
v}-\nabla\Phi-\nabla\Phi_{ext}+\sqrt{2D}{\bf R}(t).
\end{eqnarray}
The ordinary   Fokker-Planck equation associated with these equations is the mean field Kramers equation (\ref{i9}). Modeling microscopic constraints with the KIP \cite{kaniadakis}, and defining the transition probability by
\begin{eqnarray}
\label{gi3}
\pi({\bf r},{\bf v}\rightarrow {\bf v}')=w({\bf r},{\bf v},{\bf v}'-{\bf v})a\lbrack f({\bf r},{\bf v},t)\rbrack b\lbrack f({\bf r}, {\bf v}',t)\rbrack,
\end{eqnarray}
Kaniadakis \cite{kaniadakis} obtains a  nonlinear Fokker-Planck equation of the form
\begin{eqnarray}
\label{gi4} \frac{\partial f}{\partial t}+{\bf v}\cdot
\frac{\partial f}{\partial {\bf r}}-\nabla(\Phi+\Phi_{ext})\cdot \frac{\partial
f} {\partial {\bf v}} = \frac{\partial}{\partial {\bf v}}\cdot \left\lbrack \xi\left
( T h(f)\frac{\partial f}{\partial {\bf v}}+ g(f){\bf v}\right
)\right\rbrack,
\end{eqnarray}
where the functions $h(f)$ and $g(f)$ are related to the bias $a(f)$ and
$b(f)$ in the transition probabilities by
\begin{eqnarray}
\label{gi5}
g(f)=a(f)b(f), \quad h(f)=b(f)a'(f)-a(f)b'(f).
\end{eqnarray}
The generalized free energy associated with the NFP equation (\ref{gi4}) is
\begin{eqnarray}
\label{gi6}
F[f]=E-TS=\frac{1}{2}\int f v^2\, d{\bf r}d{\bf v}+\frac{1}{2}\int \rho\Phi\, d{\bf r}+\int \rho\Phi_{ext}\, d{\bf r}+T\int C(f)\, d{\bf r}d{\bf v},
\end{eqnarray}
where
\begin{eqnarray}
\label{gi7}
S=-\int C(f)\, d{\bf r}d{\bf v},\qquad C''(f)=\frac{h(f)}{g(f)},
\end{eqnarray}
is a ``generalized entropy'' determined by the ratio of the functions $h(f)$ and $g(f)$ \cite{kaniadakis,gen,frank,nfp,curado}. In the absence of microscopic constraints, $a(f)=f$ and $b(f)=1$, implying $h(f)=1$ and $g(f)=f$. In that case, Eq. (\ref{gi4}) reduces to the mean field Kramers equation (\ref{i9}), the entropy (\ref{gi7}) reduces to the Boltzmann entropy $S=-\int f\ln f\, d{\bf r}d{\bf v}$ and the free energy (\ref{gi6}) reduces to the Boltzmann free energy (\ref{e32}). We note that the NFP equation (\ref{gi4}) can be written in the form \cite{frank,nfp}:
\begin{eqnarray}
\label{gi8}
\frac{\partial f}{\partial t}+{\bf v}\cdot
\frac{\partial f}{\partial {\bf r}}-\nabla(\Phi+\Phi_{ext})\cdot \frac{\partial
f} {\partial {\bf v}}=\frac{\partial}{\partial {\bf v}}\cdot  \left ( \xi g(f)\frac{\partial}{\partial {\bf v}}\frac{\delta F}{\delta f}\right ).
\end{eqnarray}
This equation satisfies an $H$-theorem:
\begin{eqnarray}
\label{gi9}
\dot F=-\int {\xi} g(f)\left (\frac{\partial}{\partial {\bf v}}\frac{\delta F}{\delta f}\right )^2\, d{\bf r}d{\bf v}\le 0.
\end{eqnarray}
For a steady state, $\dot F=0$, Eq. (\ref{gi9}) implies that $\delta F/\delta f$ is independent on ${\bf v}$ so that the current in the NFP equation (\ref{gi8}) vanishes. Since $\partial_t f=0$, the inertial term in Eq. (\ref{gi8}) must also vanish, independently. From these two requirements, we find that $\delta F/\delta f$ is independent on ${\bf r}$. As a result,  $\delta F/\delta f$ is constant. Therefore, a steady state of the NFP equation (\ref{gi8})  is a critical point of free energy at fixed mass. Furthermore, it is dynamically stable iff it is a (local) minimum of $F$ at fixed mass $M$ (more precisely, the same results as those described at the end of Sec. \ref{sec_sza}  can be obtained). Writing $\delta F+\alpha T\delta M=0$, where $\alpha$ is a Lagrange multiplier, we find that the steady states of the generalized mean field Kramers equation (\ref{gi4}) are given by
\begin{eqnarray}
\label{gi13} f({\bf r},{\bf v})=(C')^{-1}\left\lbrace -\beta\left\lbrack \frac{v^{2}}{2}+\Phi({\bf
r})+\Phi_{ext}({\bf r})\right\rbrack-\alpha\right\rbrace.
\end{eqnarray}
Introducing the density  $\rho=\int f \, d{\bf v}$ and the kinetic pressure $p=\frac{1}{d}\int f v^2\, d{\bf v}$, and using Eq. (\ref{gi13}), we find that $\rho=\rho\lbrack \beta\Phi_{tot}({\bf r})+\alpha\rbrack$ and  $p=p\lbrack \beta\Phi_{tot}({\bf r})+\alpha\rbrack$, where $\Phi_{tot}=\Phi+\Phi_{ext}$. Eliminating $\beta\Phi_{tot}({\bf r})+\alpha$ between these two expressions, we find that the equation of state at equilibrium is barotropic: $p=p(\rho)$. We emphasize that the pressure here defined is the kinetic pressure and that the equation of state $p=p(\rho)$ is completely determined by the function $C(f)$, hence by the bias $a(f)$ and $b(f)$ in the transition probabilities. In the usual (Boltzmann) case where $C(f)=f\ln f$, we get the isothermal equation of state $p=\rho T$ (linear) but for more general functions $C(f)$, the equation of state $p(\rho)$ is nonlinear. Finally, it is easy to check (see, e.g., \cite{nfp}) that Eq. (\ref{gi13}) implies the condition of hydrostatic equilibrium
\begin{eqnarray}
\label{gi14}
\nabla p+\rho\nabla\Phi+\rho\nabla\Phi_{ext}={\bf 0}.
\end{eqnarray}
We therefore obtain the same result as in Eq. (\ref{e49}) but for a fundamentally different reason. In Eq. (\ref{e49}), the quantity $p$ designates the thermodynamical pressure $p=p_{id}+p_{ex}$, where $p_{id}=\rho k_B T/m$ is the ideal pressure (coinciding with the kinetic pressure) and  $p_{ex}$ is the excess pressure taking  into account short-range interactions. In that approach, the velocity distribution is Maxwellian leading to the isothermal gas law $p_{id}=\rho k_B T/m$ and the excess pressure comes from correlations. On the other hand, in Eq. (\ref{gi14}), the quantity $p$ designates the kinetic pressure. In that approach, correlations are ignored (or taken into account implicitly in the KIP) but the velocity distribution is non-Maxwellian leading to a nonlinear equation of state $p(\rho)$.

In the following,  we shall assume that $g(f)=f$. In that case,  the NFP equation (\ref{gi4}) becomes
\begin{eqnarray}
\label{gi10}
\frac{\partial f}{\partial t}+{\bf v}\cdot \frac{\partial f}{\partial {\bf r}}-\nabla(\Phi+\Phi_{ext})\cdot  \frac{\partial f}{\partial {\bf v}}=\frac{\partial}{\partial {\bf v}}\cdot \left\lbrack \xi \left (TfC''(f)\frac{\partial f}{\partial {\bf v}}+f{\bf v}
\right )\right\rbrack.
\end{eqnarray}
It can be derived from the generalized stochastic process
\begin{eqnarray}
\label{gi11}
{d{\bf r}\over dt}={\bf v},
\end{eqnarray}
\begin{eqnarray}
\label{gi12}
{d{\bf v}\over dt}=-\xi{\bf v}-\nabla\Phi-\nabla\Phi_{ext}+\sqrt{2Df\biggl\lbrack {C(f)\over
f}\biggr\rbrack'}{\bf R}(t),
\end{eqnarray}
where the noise explicitly depends on the distribution of particles around the particle under consideration (in phase space) \cite{borland,frank,nfp}.

\subsection{The strong friction limit}
\label{sec_gs}

In order to stress the differences with the results of Sec. \ref{sec_i}, we shall recall the derivation of the generalized Smoluchowski equation (\ref{go10})
from the generalized Kramers equation (\ref{gi10}) in the strong friction limit
$\xi\rightarrow +\infty$ \cite{nfp}. The first two moments of the
hierarchy of hydrodynamic equations associated with
Eq. (\ref{gi10}) are
\begin{eqnarray}
\label{gs1} {\partial\rho\over\partial t}+\nabla\cdot (\rho{\bf u})=0.
\end{eqnarray}
\begin{eqnarray}
\label{gs2}
\rho\left \lbrack \frac{\partial {\bf u}}{\partial t}+({\bf u}\cdot \nabla){\bf u}\right \rbrack=-\frac{\partial P_{ij}}{\partial x_{j}}-\rho\nabla\Phi-\rho\nabla\Phi_{ext}-\xi\rho {\bf u},
\end{eqnarray}
where the quantities have the same meaning as in Sec. \ref{sec_jeans}. We now consider the strong friction limit $\xi\rightarrow +\infty$
with fixed $T$. Since the term in parenthesis in Eq. (\ref{gi10})
must vanish at leading order, we find that the out-of-equilibrium
distribution function $f_0({\bf r},{\bf v},t)$ is given by
\begin{eqnarray}
\label{gs3} f_0({\bf r},{\bf v},t)=(C')^{-1}\left\lbrace -\beta\left\lbrack \frac{v^{2}}{2}+\lambda({\bf
r},t)\right\rbrack\right\rbrace+O(\xi^{-1}),
\end{eqnarray}
where $\lambda({\bf r},t)$ is a constant of integration that is
determined by the density according to
\begin{eqnarray}
\label{gs4} \rho({\bf r},t)=\int f_{0}d{\bf v}=\rho[\lambda({\bf
r},t)].
\end{eqnarray}
Note that the distribution function $f_0$ is {\it isotropic} so that
the velocity ${\bf u}({\bf r},t)=O(\xi^{-1})$ and the pressure
tensor  $P_{ij}=p\delta_{ij}+O(\xi^{-1})$ where $p$ is given by
\begin{eqnarray}
\label{gs5} p({\bf r},t)=\frac{1}{d}\int f_{0}v^{2}d{\bf
v}=p[\lambda({\bf r},t)].
\end{eqnarray}
Eliminating $\lambda({\bf r},t)$ between the two expressions
(\ref{gs4}) and (\ref{gs5}), we find that the fluid is {\it
barotropic} with an equation of state $p=p(\rho)$, the same as in equilibrium (see Sec. \ref{sec_gi}). Now, considering the momentum equation (\ref{gs2}) in
the limit $\xi\rightarrow +\infty$, we find that
\begin{eqnarray}
\label{gs6} \rho{\bf u}=-\frac{1}{\xi}(\nabla
p+\rho\nabla\Phi+\rho\nabla\Phi_{ext})+O(\xi^{-2}).
\end{eqnarray}
Inserting this relation in the continuity equation  (\ref{gs1}), we obtain
the generalized mean field Smoluchowski equation
\begin{eqnarray}
\label{gs7}\frac{\partial\rho}{\partial t}=\nabla\cdot
\left\lbrack \frac{1}{\xi}(\nabla p+\rho\nabla\Phi+\rho\nabla\Phi_{ext})\right\rbrack.
\end{eqnarray}
The free energy associated with this equation is
\begin{eqnarray}
\label{gs8} F[\rho]=\int \rho\int^{\rho}{p(\rho_1)\over
\rho_1^{2}} \,d \rho_1 d{\bf r}+{1\over 2}\int\rho\Phi d{\bf r}+\int\rho\Phi_{ext} d{\bf r}.
\end{eqnarray}
It can be deduced from the free energy (\ref{gi6}) by using
Eq. (\ref{gs3}) to express $F[f]$ as a functional $F[\rho]=F[f_0]$
of the density (see \cite{nfp} for the details of calculation). This leads to the same equations as in Sec. \ref{sec_sf} but, as explained at the end of Sec. \ref{sec_gi}, the reason is fundamentally different.

{\it Remark:} the generalized Smoluchowski equation can also be derived from the generalized Kramers equation by using a Chapman-Enskog expansion \cite{cll}. In that case, it is possible to consider generalized Kramers equations of the form (\ref{gi4}) with arbitrary  $g(f)$. This leads to generalized Smoluchowski equations of the form (\ref{gs7}) where $\xi$ now depends on position and time.

\subsection{Damped Euler equation}
\label{sec_gd}

We can also derive a damped Euler equation similar to the one obtained in Sec. \ref{sec_euler}.  To that purpose, we close the damped Jeans equation (\ref{gs2}) by using a LTE approximation \cite{gen}:
\begin{eqnarray}
\label{gd1} f_{LTE}({\bf r},{\bf v},t)=(C')^{-1}\left\lbrace -\beta\left\lbrack \frac{({\bf v}-{\bf u}({\bf r},t))^{2}}{2}+\lambda({\bf
r},t)\right\rbrack\right\rbrace,
\end{eqnarray}
where $\lambda({\bf r},t)$ is determined by the density according to
\begin{eqnarray}
\label{gd2} \rho({\bf r},t)=\int f_{LTE}d{\bf v}=\rho[\lambda({\bf
r},t)].
\end{eqnarray}
The distribution function (\ref{gd1}) minimizes the free energy (\ref{gi6}) for a given value of the density $\rho({\bf r},t)$ and local velocity ${\bf u}({\bf r},t)$. With the LTE approximation, the pressure
tensor takes the form $P_{ij}=p\delta_{ij}$ where $p$ is given by
\begin{eqnarray}
\label{gd3} p({\bf r},t)=\frac{1}{d}\int f_{LTE}({\bf v}-{\bf u}({\bf r},t))^{2}d{\bf
v}=p[\lambda({\bf r},t)].
\end{eqnarray}
Eliminating $\lambda({\bf r},t)$ between the two expressions
(\ref{gd2}) and (\ref{gd3}), we find that the fluid is
barotropic with an equation of state $p=p(\rho)$, the same as in the preceding sections. Substituting these results in Eqs. (\ref{gs1}) and (\ref{gs2}), we obtain  the damped Euler equations
\begin{eqnarray}
\label{gd4} {\partial\rho\over\partial t}+\nabla\cdot (\rho{\bf u})=0.
\end{eqnarray}
\begin{eqnarray}
\label{gd5}
\rho\left \lbrack \frac{\partial {\bf u}}{\partial t}+({\bf u}\cdot \nabla){\bf u}\right \rbrack=-\nabla p-\rho\nabla\Phi-\rho\nabla\Phi_{ext}-\xi\rho {\bf u}.
\end{eqnarray}
The free energy associated with these equations is
\begin{eqnarray}
\label{gd6} F_{tot}[\rho,{\bf u}]=\int \rho \frac{{\bf u}^2}{2}\, d{\bf r}+\int \rho\int^{\rho}{p(\rho_1)\over
\rho_1^{2}} \,d \rho_1 d{\bf r}+{1\over 2}\int\rho\Phi \, d{\bf r}+\int\rho\Phi_{ext}\,  d{\bf r}.
\end{eqnarray}
It can be deduced from the free energy (\ref{gi6}) by using
Eq. (\ref{gd1}) to express $F[f]$ as a functional $F[\rho,{\bf u}]=F[f_{LTE}]$
of the density and local velocity. This leads to the same equations as in Sec. \ref{sec_euler} but, as explained at the end of Sec. \ref{sec_gi}, the reason is fundamentally different.

\subsection{Generalized Cahn-Hilliard equations}
\label{sec_ch}

Let us assume that the long-range potential $u(|{\bf r}-{\bf r}'|)$ is screened on a distance that is large with respect to the microscopic length but short with respect to the system size. Therefore, we assume that the generalized mean field Smoluchowski equation (\ref{go9}) remains valid, but that we can simplify the potential $\Phi({\bf r},t)$ given by Eq. (\ref{o10}). Setting ${\bf q}={\bf r}'-{\bf r}$ and writing
\begin{equation}
\label{ch1}
{\Phi}({\bf r},t)=\int u(q){\rho}({\bf r}+{\bf q},t)\, d{\bf q},
\end{equation}
we Taylor expand ${\rho}({\bf r}+{\bf q},t)$ up to second order in ${\bf q}$:
\begin{equation}
\label{ch2}
{\rho}({\bf r}+{\bf q},t)={\rho}({\bf r},t)+\sum_{i}\frac{\partial{\rho}}{\partial x_{i}}q_{i}+\frac{1}{2}\sum_{i,j}\frac{\partial^{2}{\rho}}{\partial x_{i}\partial x_{j}}q_{i}q_{j}.
\end{equation}
Substituting this expansion in Eq. (\ref{ch1}), we obtain
\begin{equation}
\label{ch3}
{\Phi}({\bf r},t)=-a \rho({\bf r},t)
-\frac{b}{2}\Delta\rho({\bf r},t),
\end{equation}
with $a=-S_{d}\int_{0}^{+\infty} u(q) q^{d-1} dq$ and $b=-\frac{1}{d}
S_{d}\int_{0}^{+\infty} u(q) q^{d+1} dq$. Note that $l=(b/a)^{1/2}$
has the dimension of a length corresponding to the range of the
interaction.  Substituting Eq. (\ref{ch3}) in Eq. (\ref{go5}), we can
put the free energy in the form
\begin{eqnarray}
\label{ch5}
F\lbrack{\rho}\rbrack=\frac{b}{2}\int \left\lbrack \frac{1}{2} (\nabla{\rho})^{2}+V({\rho})\right\rbrack d{\bf r},
\end{eqnarray}
where $V$ is the effective potential
\begin{equation}
\label{ch6}
V(\rho)=-\frac{a}{b}\rho^{2}+\frac{2T}{b}C(\rho).
\end{equation}
In that case, Eq. (\ref{go7}) can be rewritten
\begin{equation}
\label{ch7}
\frac{\partial{\rho}}{\partial t}=-\nabla\cdot \left\lbrack A g({\rho})\nabla \left (\Delta{\rho}-V'({\rho})\right)\right\rbrack
\end{equation}
with $A=b/(2\xi)$. The steady state of Eq. (\ref{go7}) or (\ref{ch7})  corresponds to a uniform $\delta F/\delta\rho=-\alpha T$ leading to
\begin{equation}
\label{ch9}
\Delta{\rho}=V'({\rho})+\frac{2\alpha T}{b}.
\end{equation}
Equation (\ref{ch7}) share some analogies (but also crucial differences) with the Cahn-Hilliard equations \cite{bray}; see discussion in \cite{paper5}. Coincidentally, the case $g(\rho)=1$ and $C(\rho)=\rho^4$ gives an equation that is formally equivalent to the Cahn-Hilliard equation with $V(\rho)=\frac{2T}{b}(\frac{a}{4T}-\rho^2)^2$. On the other hand, in the classical (Boltzmann) case $g(\rho)=\rho$ and $C=\rho\ln\rho$, Eq. (\ref{ch7}) takes the form
\begin{equation}
\label{ch10}
\xi\frac{\partial{\rho}}{\partial t}=\nabla\cdot \left\lbrack (T-a\rho)\nabla \rho-\frac{b}{2}\rho\nabla (\Delta\rho) \right\rbrack,
\end{equation}
involving a density dependent diffusion coefficient $D(\rho)=\frac{1}{\xi}(T-a\rho)$ and an effective potential $\Phi_{eff}=-\frac{b}{2\xi}\Delta\rho$.

{\it Remark:} as discussed at the end of Sec. \ref{sec_w}, this gradient expansion can also be performed in the case of systems with short-range interactions described by the free energy functional (\ref{w1}), provided that we use the correspondence (\ref{w3}).

\section{Conclusion}
\label{sec_conclusion}

In this paper, we have developed a general kinetic theory of Brownian particles with long and short range interactions. To close the BBGKY-like hierarchy, we have used the  mean field approximation (\ref{zwa}) for the long-range interaction \cite{cdr} and the standard approximation (\ref{o12}) used in the theory of fluids for the short-range interaction \cite{marconi}. We have included these approximations in the general kinetic and hydrodynamic equations of Brownian particles derived in \cite{paper2,virial,paper5,hydro}. In the overdamped limit, this leads to the generalized mean field Smoluchowski equation (\ref{o14}) which includes a mean field  potential due to long-range interactions and a generically nonlinear pressure due to short-range interactions. More general equations taking into account inertial effects have also been obtained. Therefore, our kinetic theory justifies from a microscopic model the basic equations that have been introduced phenomenologically to describe various systems with long and short range interactions such as self-gravitating Brownian particles \cite{grossmann}, chemotaxis of bacterial populations \cite{ks} and colloidal particles with capillary interactions \cite{colloids}. The justification of these models from a kinetic theory starting from microscopic processes was the main goal of this paper.

We have also found that the same hydrodynamic (macroscopic) equations are obtained from nonlinear mean field Fokker-Planck equations based on generalized thermodynamics. In both cases, the nonlinear pressure takes into account microscopic constraints that affect the dynamics of the particles at small scales. However, the origin of this pressure is different. In the classical theory of fluids \cite{hansen,evans}, the distribution function is Boltzmannian and the nonlinear pressure is due to the two-body correlation function induced by the small-scale potential $u_{SR}$. Different methods have been developed in the theory of fluids to obtain the expression of the pressure law $p=p(\rho,T)$ depending on the short-range interactions. In the generalized thermodynamics approach \cite{frank,nfp,tsallisbook}, the nonlinear pressure arises from the non-Boltzmannian nature of the distribution function due to the bias in the transition probabilities from one state to the other \cite{kaniadakis}. It is interesting to observe that the hydrodynamic (macroscopic) equations coincide while the microscopic models are fundamentally different.

\appendix

\section{Expression of the free energy in the absence of strong gradients}
\label{sec_sim}

Let us consider a spatially homogeneous fluid enclosed within a container of volume $V$. The pressure $p=p(\rho,T)$ is a function of the density $\rho=M/V$ and temperature $T$. We assume that the fluid is in contact with a thermal bath imposing the temperature. Since $T$ is fixed, the pressure is barotropic so that $p=p(\rho)$. Introducing the free energy $F=E-TS$ and using the first law of thermodynamics $dE=-pdV+TdS$, we obtain the identity $dF=-pdV$ where we have used $dT=0$. This can be rewritten $dF=-pMd(1/\rho)=(p/\rho^2)Md\rho$ and, in integral form, $F=M\int^{\rho}({p}/{\rho^2})\, d\rho$. Introducing the free energy per unit volume $f=F/V$, we obtain $f(\rho)=\rho\int^{\rho}({p(\rho)}/{\rho^2})\, d\rho$. This relation remains valid {\it locally} in an inhomogeneous fluid provided that there are no strong gradients of density (e.g., the fluid is sufficiently far away from an interface). This leads to the following expression of the free energy functional
\begin{eqnarray}
\label{sim2}
F[\rho]=\int \rho\int^{\rho}\frac{p(\rho_1)}{\rho_1^2}\, d\rho_1.
\end{eqnarray}

\section{Some comments about the importance of metastable states}
\label{sec_comments}

In this Appendix, we discuss the importance of metastable states in the kinetic theory of systems with long (and short) range interactions. This will help us to better understand the relation between deterministic and stochastic kinetic equations.

As emphasized by Archer \& Rauscher \cite{archer}, we must distinguish three types of density fields: (i) the density operator $\rho_{d}({\bf r},t)=\sum_{i=1}^N m\delta({\bf r}_i(t)-{\bf r})$ which is made of a sum of Dirac peaks coinciding with the exact positions of the particles, (ii) the ensemble average density field $\rho({\bf r},t)=\langle \sum_{i=1}^N m\delta({\bf r}_i(t)-{\bf r})\rangle =NmP_1({\bf r},t)$, and (iii) the coarse-grained density field $\overline{\rho}({\bf r},t)$ which can be viewed either as a spatial \cite{kawasaki} or time \cite{archer} average of the density operator $\rho_{d}({\bf r},t)=\sum_{i=1}^N m\delta({\bf r}_i(t)-{\bf r})$. Note that this distinction is standard in equilibrium statistical mechanics. Historically, it first appeared in Boltzmann's combinatorial analysis. A microstate is characterized by the density $\rho_{d}({\bf r},t)$ specifying the exact position of all the particles while a   macrostate is characterized by the coarse-grained  density  $\overline{\rho}({\bf r})$ such that $\overline{\rho}({\bf r})d{\bf r}$ gives the number of particles in a macrocell $[x,x+dx]\times [y,y+dy]\times [z,z+dz]$ irrespectively of the exact positions of the particles in the cell.  At statistical equilibrium, the density probability of the coarse-grained density field $\overline{\rho}({\bf r})$ is
\begin{equation}
\label{meta1}
P_{eq}[\overline{\rho}]=\frac{1}{Z(\beta)} e^{-\beta F[\overline{\rho}]}\delta(M-M[\overline{\rho}]),
\end{equation}
where $F[\overline{\rho}]$ is the free energy (\ref{e33}). The normalization condition  $\int P_{eq}[\overline{\rho}]\, {\cal D}\overline{\rho}=1$ leads to the expression of the partition function $Z(\beta)=\int e^{-\beta F[\overline{\rho}]}\delta(M-M[\overline{\rho}])\, {\cal D}\overline{\rho}$. The ensemble average density $\rho_{eq}({\bf r})$ corresponds to the average value of $\overline{\rho}({\bf r})$, i.e. ${\rho}_{eq}({\bf r})=\int P_{eq}[\overline{\rho}]\overline{\rho}\, {\cal D}\overline{\rho}$. At the thermodynamic limit, it coincides with the most probable value of $\overline{\rho}({\bf r})$. Therefore,  ${\rho}_{eq}({\bf r})$ corresponds to the global minimum ${{\rho}}_{global}({\bf r})$ of $F[\overline{\rho}]$. For systems with long-range interactions, this has been proven rigorously  in \cite{messer}.

When the free energy $F[\overline{\rho}]$ has a unique (global) minimum, the situation is simple. The density probability $P_{eq}[\overline{\rho}]$ obtained at a given time $t$ from an  ensemble of experiments, or the density probability $P_{eq}[\overline{\rho}]$ obtained from the time series of a  unique experiment, coincide and are given by Eq. (\ref{meta1}).  On the other hand, the average (or most probable) value of $\overline{\rho}({\bf r})$ is given by $\rho_{eq}({\bf r})$  which is the (global) minimum of $F[\overline{\rho}]$.  Let us now consider the more complicated situation where $F[\overline{\rho}]$ has several minima (metastable states). This occurs in particular for systems with long-range interactions, like self-gravitating systems \cite{metastable,ijmpb},  and we shall focus on these systems in the following discussion. If we fix the time $t$ large enough \footnote{We will see that $t$ has to be large with respect to $e^N$, so that it has to be {\it very} large in practice!} and consider an ensemble of experiments, we will measure a coarse-grained density field $\overline{\rho}({\bf r})$ that fluctuates from experiment to experiment. Its density distribution will be given by Eq. (\ref{meta1}). The system will be found most of the time in a minimum of $F[\overline{\rho}]$, the global minimum being the most frequent one. Recalling that for systems with long-range interactions the free energy is extensive, the equilibrium density probability of the coarse-grained distribution (\ref{meta1}) can be rewritten
\begin{equation}
P_{eq}[\overline{\rho}]=\frac{1}{Z(\beta)} e^{-\beta N f[\overline{\rho}]}\delta(M-M[\overline{\rho}]),
\end{equation}
where $f[\overline{\rho}]=F[\overline{\rho}]/N$ is independent on $N$. For $N\rightarrow +\infty$, the distribution is strongly peaked around the global minimum of $F[\overline{\rho}]$ at fixed mass, so that an overwhelming majority of configurations with $\overline{\rho}({\bf r})\simeq {\rho}_{global}({\bf r})$ we will observed.  Accordingly, the partition function is dominated by the contribution of the global minimum and we can make the approximation $Z(\beta)\simeq e^{-\beta N f_{min}(\beta)}$ so that  $F(\beta)=-\frac{1}{\beta}\ln Z(\beta)\simeq N f_{min}(\beta)$ where $f_{min}(\beta)=f[\rho_{global}]$. Equivalently, we have $P_{eq}[\overline{\rho}]\simeq \delta(\overline{\rho}-\rho_{global})$. However, focussing exclusively on the distribution $P_{eq}[\overline{\rho}]$ and on the ensemble average $\rho_{eq}({\bf r})$ may hide the importance of metastable states in the dynamics \footnote{This is particularly true for self-gravitating systems for which there is no global minimum of free energy due to gravitational
collapse. Yet, the system can be found in a metastable state (local
minimum of free energy) that can persist for a very long time.}. To see that, let us now consider one experiment and follow the system in time. If $N$ is not too large and/or if we wait long enough, we will see that the system undergoes random transitions from one metastable state to the other.  Of course, the global
minimum of free energy is the most frequently visited. The residence time in a metastable state depends on the barrier of free energy with the other minima and is given by the Kramers formula $t_{life}\sim e^{\beta|\Delta F|}$. If we compute  the pdf of $\overline{\rho}({\bf r},t)$ on a time series {\it over sufficiently long times}, we will obtain the distribution (\ref{meta1}). However, we insist on the fact that the time on which we calculate the pdf must be extremely long (recall that equilibrium statistical mechanics assume ideally  that $t\rightarrow +\infty$).  Indeed, since the free energy is extensive, the barrier of free energy  scales linearly with the number of particles ($|\Delta F|\sim N$) and the lifetime of a metastable state scales like  $t_{life}\sim e^N$ \cite{metastable}. Therefore, when $N$ is large (it has not to be {\it very} large since the dependence of $t_{life}$ with $N$ is exponential), the metastable states will have tremendously long lifetimes! In practice, the system may remain blocked in a metastable state $\overline{\rho}({\bf r},t)\simeq \rho_{meta}({\bf r})$ for all the duration of the physical experiment.  In that case, we will  measure $P_{phys}[\overline{\rho}]\simeq \delta(\overline{\rho}-\rho_{meta})$ instead of $P_{eq}[\overline{\rho}]$. This leads to an apparently non ergodic behavior, although ergodicity holds provided that we wait long enough. These arguments show that the limits $N\rightarrow +\infty$ and $t\rightarrow +\infty$ do not commute. If we fix $N$ and make an experiment on a duration $t\rightarrow +\infty$, we will obtain $P_{eq}[\overline{\rho}]$ given by Eq. (\ref{meta1}) and $\rho({\bf r})=\rho_{global}({\bf r})$. Alternatively, if we fix an interval of time $[0,T]$ that is large but finite and let $N\rightarrow +\infty$, the system may remain blocked in a metastable state during all the duration of the experiment (since its lifetime diverges like $e^N\rightarrow +\infty$). In that case, we will find $P_{phys}[\overline{\rho}]\simeq \delta(\overline{\rho}-\rho_{meta})$ and $\rho({\bf r})=\rho_{meta}({\bf r})$. Having realized that, we can now better understand the relation between the deterministic and the stochastic kinetic equations.

The time evolution of the density operator $\rho_{d}({\bf r},t)=\sum_{i=1}^N m\delta({\bf r}_i(t)-{\bf r})$ is given by the exact stochastic kinetic equation  \cite{dean}:
\begin{eqnarray}
\label{meta2}
\frac{\partial {\rho}_d}{\partial t}=\nabla\cdot \left ( \frac{1}{\xi}{\rho}_d\nabla\frac{\delta F_d}{\delta {\rho}_d}\right )+\nabla\cdot \left (\sqrt{\frac{2k_{B}T {\rho}_d}{\xi}}{\bf R}\right ),
\end{eqnarray}
where
\begin{eqnarray}
\label{meta3}
F_d[\rho_d]=\frac{1}{2}\int \rho_d({\bf r},t) u(|{\bf r}-{\bf r}'|)\rho_d({\bf r}',t)\, d{\bf r}d{\bf r}'+\int \rho_d \Phi_{ext}\, d{\bf r}+T\int \frac{\rho_d}{m}\ln \frac{\rho_d}{m}\, d{\bf r}d{\bf r}',
\end{eqnarray}
is the exact free energy. This equation bears exactly the same information as the $N$-body dynamics (\ref{o1}) and, as such, is not very useful for practical applications. The stochastic kinetic equation (\ref{meta2}) can be viewed as a Langevin equation with a multiplicative noise that vanishes when $\rho_d=0$. This expresses the fact that the density cannot fluctuate in regions devoid of particles. The corresponding Fokker-Planck equation for the density probability $P[\rho_d,t]$ of the distribution $\rho_{d}({\bf r},t)$ is given by \cite{frusawa}:
\begin{eqnarray}
\label{meta4}
\xi\frac{\partial P[\rho_d,t]}{\partial t}=-\int \frac{\delta}{\delta\rho_d({\bf r},t)}\left\lbrace \nabla\cdot \rho_d({\bf r},t)\nabla\left\lbrack T\frac{\delta}{\delta\rho_d({\bf r},t)}+\frac{\delta F_d}{\delta\rho_d({\bf r},t)}\right\rbrack P[\rho_d,t]\right\rbrace\, d{\bf r}.
\end{eqnarray}
The steady state of this Fokker-Planck equation is $P_{eq}[\rho_d]\propto e^{-\beta F_d[\rho_d]}\delta(\int\rho_d\, d{\bf r}-N)$ \cite{dean} which is equivalent to the canonical $N$ body distribution (\ref{e6}). The  ensemble average density
$\rho({\bf r},t)=\langle \sum_{i=1}^N m\delta({\bf r}_i(t)-{\bf
r})\rangle$, which is a  deterministic field, satisfies the exact Smoluchowski equation (\ref{o7}). It can be obtained by averaging the exact stochastic equation for  $\rho_{d}({\bf r},t)$ \cite{marconi,paper5} or by writing  the first exact equation of the BBGKY hierarchy \cite{paper2,archerevans}. However, this equation is not closed and some approximations, whose validity will be discussed below,  must be introduced.
Finally, the coarse-grained density $\overline{\rho}({\bf r},t)$ is a fluctuating field whose
evolution is governed by the stochastic kinetic equation  (\ref{s6}) with the free energy (\ref{e33}). This equation can be obtained in a strong friction limit of fluctuating hydrodynamic equations \cite{munakata}, by coarse-graining the exact stochastic equation (\ref{meta2}) \cite{archer} or by using the general theory of fluctuations of Landau \& Lifshitz \cite{ll} (see Appendix B of \cite{paper5}). The stochastic kinetic equation (\ref{s6}) can be viewed as a Langevin equation for the coarse-grained density $\overline{\rho}({\bf r},t)$. The density probability $P[\overline{\rho},t]$ of the coarse-grained density $\overline{\rho}({\bf r},t)$ is given by the Fokker-Planck equation \cite{munakata,kawasaki}:
\begin{eqnarray}
\label{meta5}
\xi\frac{\partial P[\overline{\rho},t]}{\partial t}=-\int \frac{\delta}{\delta\overline{\rho}({\bf r},t)}\left\lbrace \nabla\cdot \overline{\rho}({\bf r},t)\nabla\left\lbrack T\frac{\delta}{\delta\overline{\rho}({\bf r},t)}+\frac{\delta F}{\delta\overline{\rho}({\bf r},t)}\right\rbrack P[\overline{\rho},t]\right\rbrace\, d{\bf r},
\end{eqnarray}
where $F[\overline{\rho}]$ is the free energy (\ref{e33}).  The steady state of this Fokker-Planck equation  is the statistical equilibrium state (\ref{meta1}) \cite{munakata,kawasaki}. The convergence towards this equilibrium state is guaranteed by an $H$-theorem \cite{munakata1994}. The density $\rho({\bf r},t)$ is the  average value of $\overline{\rho}({\bf r},t)$, i.e. ${\rho}({\bf r},t)=\int P[\overline{\rho},t]\overline{\rho}\, {\cal D}\overline{\rho}$. In the thermodynamic limit, it coincides with the most probable value of $\overline{\rho}({\bf r},t)$. As we have previously indicated, the density $\rho({\bf r},t)$ satisfies the exact Smoluchowski equation (\ref{o7}), but this equation is not closed. The stochastic kinetic equation (\ref{s6}), which is closed (unlike the exact Smoluchowski equation (\ref{o7})) and which describes the evolution of a smooth field (unlike the exact stochastic equation (\ref{meta2}) which describes the evolution of an operator made of Dirac peaks) is essentially ``exact'' and represents the most important equation of the list.

Let us now discuss the validity of the approximate  Smoluchowski equation (\ref{o15}) in the light of the previous considerations. If the functional $F[\overline{\rho}]$ has a unique (global) minimum, then the solution of the deterministic  equation (\ref{o15}) converges towards this minimum for $t\rightarrow +\infty$.  Therefore, the density $\rho({\bf r},t)$ tends to the equilibrium density $\rho_{eq}({\bf r})$ and Eq. (\ref{o15}) certainly provides a good description of the average dynamics. However, when the free energy functional $F[\overline{\rho}]$ possesses several  local minima (metastable states), the situation is more complicated. It that case, the deterministic kinetic equation (\ref{o15}) will converge for $t\rightarrow +\infty$ to one of these minima (local or global), whose selection will depend on a notion of basin of attraction.  Since Eq. (\ref{o15}) is a deterministic equation, the system  will remain in that state $\rho_{meta}({\bf r},t)$ for ever, even if this is not the global minimum of $F[\overline{\rho}]$. Therefore, different initial conditions (belonging to different basins of attraction) will lead to different density profiles $\rho({\bf r})$ for $t\rightarrow +\infty$ in contradiction with the fact that $\rho({\bf r},t)$ should tend to a unique profile $\rho_{eq}({\bf r})$ which is the average (or most probable) value of $\overline{\rho}({\bf r})$ according to the distribution (\ref{meta1}). This indicates that the  approximate deterministic equation (\ref{o15}) is not correct when there exists  metastable states since it may not converge towards the correct equilibrium state. By contrast, the average (or most probable) density profile $\rho({\bf r},t)$ determined from the solution of the stochastic equation (\ref{s6}) correctly tends towards the  equilibrium profile  $\rho_{eq}({\bf r})$ for $t\rightarrow +\infty$. This shows that the stochastic equation (\ref{s6}) is superior to the deterministic equation (\ref{o15}).

However, for systems with long-range interactions, it has been proven rigorously that the mean field approximation is exact for $N\rightarrow +\infty$, so that the mean field Smoluchowski equation (\ref{o9}) should be exact in that limit even if there exists metastable states. How can we solve this apparent paradox? The solution comes from the non-commutation of the limits $t\rightarrow +\infty$ and $N\rightarrow +\infty$ that we have previously indicated \footnote{Note that the non-commutation of the limits that we consider here is different from the one that occurs in relation to quasi stationary states (QSS) in the collisionless regime of systems with long-range interactions \cite{latora,incomplete}. These authors consider an isolated Hamiltonian system and discuss the difference between QSSs (which are steady states of the Vlasov equation) and the statistical equilibrium state (global entropy maximum). Here, we consider an overdamped Brownian system and discuss the difference between metastable states (local free energy minima) and the strict equilibrium state (global free energy minimum). Note that our discussion on the importance of metastable states could also apply to isolated Hamiltonian systems with long-range interactions when the Boltzmann entropy $S[f]$ has several maxima at fixed mass and energy. In that case, the system can achieve a Vlasov QSS on a timescale of order $1$, then exhibit random changes between different microcanonical metastable states. These  microcanonical metastable states appear on a typical timescale $t_{relax}(N)$ diverging with $N$ and their lifetime scales like $e^N$ \cite{metastable}. If the system is in contact with a heat bath \cite{baldovin}, and if the Boltzmann free energy $F[f]$ has several minima at fixed mass, the system will ultimately exhibit random changes between canonical metastable states. These  canonical metastable states appear on a typical timescale $\sim 1/\xi$ \cite{baldovin} and their lifetime scales like $e^N$ \cite{metastable}. In that situation, the Vlasov steady states (formed on a timescale $1$) and the microcanonical equilibrium states (formed on a timescale $t_{relax}(N)$) are {\it quasistationary} and can be destroyed by the effect of the thermal bath \cite{baldovin}. Therefore, depending on the values of $t$, $N$ and $\xi$, a rich variety of dynamical behaviors can occur in systems with long-range interactions.}. The validity of the mean field Smoluchowski equation (\ref{o9}) assumes that we fix the interval of time $[0,T]$ (any), then let $N\rightarrow +\infty$. In that case, we have seen that the lifetime of a metastable state tends to $e^N\rightarrow +\infty$ so that it is larger than the physical interval $[0,T]$. This is the reason why the  solution of the mean field Smoluchowski equation can generically converge towards {\it any} minimum (local or global) of $F[\rho]$ and stay there permanently (it does not make any difference between fully
stable or metastable states). Alternatively, if we fix $N$ (any), then let $t\rightarrow +\infty$, the evolution of $\rho({\bf r},t)$ is {\it not} described by the mean field Smoluchowski equation. In that case, the coarse-grained density $\overline{\rho}({\bf r},t)$ undergoes random changes from one metastable state to the other and we must consider the stochastic equation (\ref{s6}) if we want to take into account these random changes properly.

In practice, if $N$ is large, the system may be trapped for all physically relevant times in a metastable state which is not the global free energy minimum. Still, this metastable state is fully relevant on a physical point of view. It has an extremely long lifetime,
scaling like $e^N$ \cite{metastable}, so that it is as much relevant as
the fully stable state when $N\rightarrow +\infty$. This is a situation of physical ergodicity breaking (although, strictly speaking, the system is ergodic if we wait long enough). In that case, the system will not jump to another  metastable state in the duration of the experiment, so that  we can use the mean field  Smoluchowski equation (\ref{o9})  which is deterministic. On the other hand, if $N$ is not too large, or if we are close to a critical point so that the barrier of free energy $|\Delta F|$ is small \cite{paper5}, we will observe random changes from one metastable state to the other in the duration of the experiment. In that case, we must use the stochastic kinetic equation (\ref{s6}). Typically, we must use the stochastic kinetic equation (\ref{s6}) when the physical timescale of the experiment is larger that $e^N$ or more precisely $e^{N|\Delta f|}$. Therefore, the domain of validity of the mean field Smoluchowski equation (\ref{o9}) for systems with long-range interactions is clearly established. By contrast, the domain of validity of the approximate equation (\ref{o15}) for systems with short range interactions remains less clear when there exists metastable states since there is no small parameter (like $1/N$ in the previous case) in the limit of which this equation can be rigorously justified. Nevertheless, by analogy, we can argue that this equation can be employed when the timescale of the experiment is much smaller than the typical lifetime of a metastable state.

\end{document}